\documentclass[12pt]{article} 

\usepackage{graphicx}
\usepackage[a4paper, total={7in, 8in}]{geometry}

\usepackage{amsmath}
\usepackage{booktabs}
\usepackage{setspace}
\usepackage{amsfonts}
\usepackage{amssymb}
\usepackage{amsthm} 
\usepackage{multicol}
\usepackage{graphicx}
\usepackage{enumerate}
\usepackage{xspace}
\usepackage[inline,shortlabels]{enumitem}
\usepackage{bbm}
\usepackage{bm}
\usepackage[usenames,dvipsnames]{color}
\usepackage[colorlinks=true,linkcolor=blue, citecolor=blue, urlcolor=Salmon]{hyperref}
\usepackage{fancyhdr}
\usepackage{floatpag}

\def\vec#1{{\bm #1}}

\def\vec#1{{\bm #1}}

\def\mat#1{\mathbf{#1}}

\def\given{\mid}

\usepackage{url}
\usepackage[normalem]{ulem}

\usepackage{color}
\usepackage[dvipsnames]{xcolor}

\def\del#1{}

\usepackage{booktabs}
\usepackage{authblk}
\usepackage{setspace}

\def\vec#1{{\bm #1}}

\def\vec#1{{\bm #1}}

\def\mat#1{\mathbf{#1}}

\def\given{\mid}

\usepackage{url}
\usepackage{ulem}
\usepackage{color}

\usepackage{multirow}
\usepackage{booktabs}
\usepackage{authblk}
\usepackage{setspace}

\DeclareEmphSequence{\itshape}
\title{Network community detection via neural embeddings}

\author[1,2]{Sadamori Kojaku}
\author[1]{Filippo Radicchi}
\author[1]{Yong-Yeol Ahn}
\author[1,3]{Santo Fortunato}

\affil[1]{Center for Complex Networks and Systems Research, Luddy School of Informatics, Computing, and Engineering, Indiana University, Bloomington, USA}
\affil[2]{School of Systems Science and Industrial Engineering, Binghamton University, Binghamton, New York USA}
\affil[3]{santo.fortunato@indiana.edu}
\date{}

\begin{document}

\maketitle

\begin{abstract}
    Recent advances in machine learning research have produced powerful neural graph embedding methods, which learn useful, low-dimensional vector representations of network data.
    These neural methods for graph embedding excel in graph machine learning tasks and are now widely adopted.
    However, how and why these methods work---particularly how network structure gets encoded in the embedding---remain largely unexplained.
    Here, we show that node2vec---shallow, linear neural network---encodes communities into separable clusters better than random partitioning down to the information-theoretic detectability limit for the stochastic block models.
    We show that this is due to the equivalence between the embedding learned by node2vec and the spectral embedding via the eigenvectors of the symmetric normalized Laplacian matrix.
    Numerical simulations demonstrate that node2vec is capable of learning communities on sparse graphs generated by the stochastic blockmodel, as well as on sparse degree-heterogeneous networks.
    Our results highlight the features of graph neural networks that enable them to separate communities in embedding space.
\end{abstract}


\section{Introduction}

Networks represent the structure of complex systems as sets of nodes connected by edges~\cite{barabasiNetworkScience2016,menczerFirstCourseNetwork2020,newmanNetworks2018} and are ubiquitous across diverse domains, including social sciences~\cite{misloveUnderstandingDemographicsTwitter2011,kojakuEffectivenessBackwardContact2021}, transportation~\cite{barthelemySpatialNetworks2011,colizzaRoleAirlineTransportation2006}, finance~\cite{bardosciaPhysicsFinancialNetworks2021,baruccaNetworkValuationFinancial2020}, science of science~\cite{newmanStructureScientificCollaboration2001,clausetSystematicInequalityHierarchy2015}, neuroscience~\cite{bassettNetworkNeuroscience2017,SmallWorldBrainNetworks}, and biology~\cite{Kim2013DiscoveryOA,Samaga2013ModelingAF,Rozum2020ParityAT}.
Networks are complex, high-dimensional, and discrete objects, making it highly non-trivial to obtain useful representations of their structure.
For instance, recommendation systems for social networks typically require informative variables (or ``features'') that capture the most important structural characteristics. Often, these features are designed through trial and error, and may not be generalizable across networks.

Graph embeddings automatically identify useful structural features for network elements, most commonly for the nodes~\cite{vonluxburgTutorialSpectralClustering2007,kunegisLearningSpectralGraph2009}.
Each node is represented as a point in a compact and continuous vector space.
Such a vector representation enables the direct application of
powerful machine learning methods, capable of solving various tasks, such as visualization~\cite{nickelPoincareEmbeddingsLearning2017,pengNeuralEmbeddingsScholarly2021}, clustering~\cite{barotCommunityDetectionUsing2021,tandonCommunityDetectionNetworks2021}, and prediction~\cite{chenPMEProjectedMetric2018,kunegisLearningSpectralGraph2009,masrourBurstingFilterBubble2020}.
This representation can facilitate the operationalization of abstract concepts using vectorial operations~\cite{kwakFrameAxisCharacterizingMicroframe2021,murrayUnsupervisedEmbeddingTrajectories2021,pengNeuralEmbeddingsScholarly2021,souratiAcceleratingScienceHuman2021,tshitoyanUnsupervisedWordEmbeddings2019}.
Graph embeddings have been studied in various contexts.
For example, spectral embedding stems from the spectral analysis of networks~\cite{vonluxburgTutorialSpectralClustering2007,newmanFindingCommunityStructure2006}. A closely related formulation is matrix factorization~\cite{qiuNetworkEmbeddingMatrix2018,penningtonGloVeGlobalVectors2014}.
Recent years have witnessed a substantial shift towards a new paradigm of graph embeddings based on neural networks~\cite{agarwalUnifiedFrameworkFair2021,dehghan-kooshkghaziEvaluatingNodeEmbeddings2022,groverNode2vecScalableFeature2016,hamiltonInductiveRepresentationLearning2017,liuUnderstandingOnsetHot2021,mengAnalysisNode2vecRandom2020,pengNeuralEmbeddingsScholarly2021,perozziDeepWalkOnlineLearning2014,tandonCommunityDetectionNetworks2021,tangLINELargescaleInformation2015,velickovicGraphAttentionNetworks2022}, which have demonstrated remarkable effectiveness across many computational tasks~\cite{chenPMEProjectedMetric2018,groverNode2vecScalableFeature2016,perozziDeepWalkOnlineLearning2014,tangLINELargescaleInformation2015,hamiltonInductiveRepresentationLearning2017,tangLINELargescaleInformation2015,velickovicGraphAttentionNetworks2022}.
Yet, due to the inherent black-box nature of neural networks, how and why these methods work is still largely unknown, and we lack a clear understanding of the process of encoding certain network structure onto embeddings.

One of the fundamental and ubiquitous features of networks is community structure, i.e., the existence of cohesive groups of nodes,
characterized by a density of within-group edges that is
higher than the density of edges between them~\cite{fortunatoCommunityDetectionGraphs2010,fortunatoCommunityDetectionNetworks2016,fortunato20YearsNetwork2022}.
In practice, neural graph embedding methods are widely used to discover communities from networks~\cite{penningtonGloVeGlobalVectors2014,groverNode2vecScalableFeature2016,perozziDeepWalkOnlineLearning2014,murrayUnsupervisedEmbeddingTrajectories2021}.

The stochastic block model (SBM) is a basic generative model of networks with community structure ~\cite{peixotoParsimoniousModuleInference2013,karrerStochasticBlockmodelsCommunity2011} and is regularly used as a benchmark for community detection algorithms.
Some community detection methods are able to correctly classify all nodes into communities in large and dense networks generated by the SBM, provided that the average degree increases as the number of nodes increases~\cite{zhangExactRecoveryCommunity2022,chenUniversalPhaseTransition2015,chenPhaseTransitionsSpectral2015,barotCommunityDetectionUsing2021,zhangConsistencyRandomwalkBased2021,Abbe2015CommunityDI}.
However, most networks of interest in applications are sparse~\cite{yangNetworksIntroductionNewman2013,barabasiNetworkScience2016}, in that their average degree is usually much smaller than the network size.  The task of community detection is particularly hard on very sparse networks.
For instance, the performance of many spectral methods significantly worsens as the graph gets sparser ~\cite{krzakalaSpectralRedemptionClustering2013,benaych-georgesSpectralRadiiSparse2021}, which has led to the development of remedies such as non-backtracking walks~\cite{krzakalaSpectralRedemptionClustering2013,newmanSpectralCommunityDetection2013,benaych-georgesSpectralRadiiSparse2021} and consensus clustering~\cite{zhangScalableDetectionStatistically2014}.
However, it remains unclear how neural graph embeddings perform on sparse networks, how much edge sparsity hampers their ability to detect communities, and how they fare for traditional clustering techniques, especially spectral methods.

Here, we prove that
graph embedding methods based on a shallow neural network without non-linear activation---such as DeepWalk~\cite{perozziDeepWalkOnlineLearning2014}, LINE~\cite{tangLINELargescaleInformation2015}, and node2vec~\cite{groverNode2vecScalableFeature2016}---can resolve
communities all the way down to the information-theoretical limit on graphs generated by the SBM~\cite{decelleAsymptoticAnalysisStochastic2011}.
Our results imply that two common components of deep learning---multiple deep layers and non-linear activation---are not necessary to achieve the optimal limit of community detectability.
Numerical experiments demonstrate that the communities embedded by node2vec can be effectively identified by the $K$-means algorithm, with accuracy close to the performance of the optimal belief propagation (BP) method~\cite{decelleAsymptoticAnalysisStochastic2011} when the true number of communities is given to the $K$-means and BP.
Additional numerical tests reveal that node2vec is able to learn communities also in the presence of heterogeneity of degree and community size.
In this case the two-step approach, combining embedding and clustering, is underperforming in certain settings, but this may be due to the fact that $K$-means clustering struggles when clusters have widely different sizes. We expect that addressing this shortcoming of $K$-means would lead to much better results.

Our work might help to inform powerful community detection algorithms and improve our theoretical understanding of clustering via neural embeddings. The code to reproduce all the results is available at \cite{github}.

\section{Results}\label{sec:result}

\subsection{Planted Partition model}
\label{sec:ppm}

We first consider the standard setting studied in papers concerning community detectability~\cite{benaych-georgesSpectralRadiiSparse2021,krzakalaSpectralRedemptionClustering2013,nadakuditiGraphSpectraDetectability2012}.
We focus on undirected and unweighted networks with community structure generated according to the planted partition model (PPM)~\cite{condonAlgorithmsGraphPartitioning1999}, a special case of the SBM where nodes are divided into $q$ equal-sized communities, and two nodes are connected with probability $p_{\text{in}}$ if they are in the same community and with probability $p_{\text{out}}$ if they are in different communities.
We assume that the networks are sparse, i.e.,  $p_{\text{in}}$ and $p_{\text{out}}$ are inversely proportional to the number $n$ of nodes.
The average degree $\langle k \rangle$ and the ratio of edge probabilities $p_{\text{in}} / p_{\text{out}}$ do not depend on $n$.
We specify the edge probabilities via the mixing parameter $\mu = n p_{\text{out}} / \langle k \rangle$.
The mixing parameter indicates how blended communities are with each other. As $\mu \rightarrow 0$,  communities are well separated and easily detectable. For larger values of $\mu$, community detection becomes harder. For $\mu=1$, which corresponds to $p_{\text{in}}= p_{\text{out}}$, the network is an Erd\H{o}s-R\'enyi random graph and, as such, has no community structure.
We note that the mixing parameter $\mu$ is slightly different from the traditional mixing parameter $\mu_{\text{LFR}}$ used in the Lancichinetti-Fortunato-Radicchi (LFR) benchmark, which is defined as $\mu_{\text{LFR}} = (1-\frac{1}{q})n p_{\text{out}} / \langle k \rangle$. The difference between $\mu$ and $\mu_{\text{LFR}}$ is negligible for large $q$.

\subsection{Detectability limit of communities}
\label{sec:detectability_limit}
The goal of community detection in the PPM is to recover the block membership of the model based on the structure of the specific networks generated by it. When communities are well separated, an algorithm is likely to recover these communities perfectly. However, as the number of inter-community edges increases, thereby reducing the difference between the densities of inter-community and intra-community edges, the algorithm may fail to correctly classify some nodes, and eventually, communities cannot be detected better than random guessing.
The level of community mixing above which no algorithm can recover communities better than random guessing is the information-theoretic detectability limit~\cite{nadakuditiGraphSpectraDetectability2012,decelleAsymptoticAnalysisStochastic2011}.

Operationally, with the PPM, the level of mixing is quantified by $\mu$. Communities are present for all $\mu$-values in the range $[0, 1)$, because the edges are more densely distributed within communities than between them.
In the regime above the information-theoretical limit (i.e., $\mu^* \leq \mu < 1$), communities are not detectable because their inter-community/intra-community edge densities are indistinguishable from the corresponding edge densities of random partitions.

\subsection{Detectability limit of node2vec}

We first give a high-level description of our derivation of the algorithmic detectability limit for node2vec. We note that our derivation can be directly applied to other neural graph embeddings such as DeepWalk~\cite{perozziDeepWalkOnlineLearning2014} and LINE~\cite{tangLINELargescaleInformation2015}.
See the Methods section for the step-by-step derivations.

Our analysis is based on the fact that node2vec generates its embedding by effectively factorizing a matrix when the number of dimensions is sufficiently large~\cite{qiuNetworkEmbeddingMatrix2018}. This insight enables us to study node2vec as a spectral method (see Methods).
Spectral algorithms identify communities by computing the eigenvectors associated with the largest or smallest eigenvalues of a reference operator such as the combinatorial and normalized Laplacian matrices.
When using eigenvectors to represent the network in vector space, nodes in the same community are projected onto points in space lying close to each other so that a data clustering algorithm can separate them~\cite{vonluxburgTutorialSpectralClustering2007}.

The existence of such localized eigenvectors can be inferred by
analyzing the spectrum of the reference operator using random matrix theory.
For instance, this approach has been applied to determine the detectability limit of the normalized Laplacian matrix generated by the PPM ~\cite{RadicchiDetectabilityHeterogeneousNetworks2013}. We find that, under some mild conditions, the spectrum of the node2vec matrix is equivalent to that of the normalized Laplacian matrix. Hence, the detectability limit of node2vec matches that of the spectral embedding with the normalized Laplacian matrix~\cite{RadicchiDetectabilityHeterogeneousNetworks2013}:
\begin{equation}
    \mu^*_{\text{n2v}} = \mu^* = 1 - \frac{1}{\sqrt{\langle k \rangle}}.
    \label{eq:detectability_limit}
\end{equation}
See Supporting Information Section 2 for the expression of the detectability limit in terms of the mixing parameter $\mu$.
This threshold exactly corresponds to the information-theoretical detectability limit $\mu^*$ of the PPM~\cite{nadakuditiGraphSpectraDetectability2012,zhangScalableDetectionStatistically2014}.
In other words, node2vec has the ability to detect communities down to the information-theoretic limit in principle.
However, like in the case of spectral modularity maximization~\cite{nadakuditiGraphSpectraDetectability2012}, our analysis is only valid when the average degree is sufficiently large. Nevertheless, as we shall see, our numerical simulations show that node2vec performs well even if the average degree is small.

\subsection{Experiment setup}

As baselines, we use two spectral embedding methods whose detectability limit matches the information-theoretical one: spectral modularity maximization~\cite{nadakuditiGraphSpectraDetectability2012} and Laplacian EigenMap~\cite{belkinLaplacianEigenmapsDimensionality2003}. In addition, we use two other neural embeddings, DeepWalk~\cite{perozziDeepWalkOnlineLearning2014} and LINE~\cite{tangLINELargescaleInformation2015}.
DeepWalk and LINE share the same architecture as node2vec but are trained with different objective functions~\cite{qiuNetworkEmbeddingMatrix2018,kojakuResidual2VecDebiasingGraph2021}.
Furthermore, we employ the spectral algorithm based on the leading eigenvectors of the non-backtracking matrix, which reaches the information-theoretical limit even in the sparse case for networks generated by the PPM  ~\cite{krzakalaSpectralRedemptionClustering2013}.
For all embedding methods, we set the number of dimensions, $C$, to $64$.
Finally, we employ two community detection algorithms: statistical inference of the microscopic degree-corrected SBM~\cite{peixotoParsimoniousModuleInference2013}, and the BP algorithm~\cite{decelleAsymptoticAnalysisStochastic2011}.
The BP algorithm is theoretically optimal for PPM networks and serves as an ideal baseline for assessing graph embeddings. However, achieving optimal performance with BP in practice requires non-trivial parameter tuning.
Therefore, we initialized the BP using the information about the true communities, namely the number of nodes in each true community and the number of edges between the communities.
See Supporting Information Section 4 for the parameter choices of the models and the implementations we used.

Community detection via graph embedding is a two-step process:
\begin{itemize}
    \item First, the network is embedded, which yields a projection of nodes onto points in a vector space.
    \item Second, the points are divided into groups using a data clustering method (e.g., $K$-means clustering).
\end{itemize}

Thus, the performance of community detection depends on both the quality of the embedding and the performance of the subsequent data clustering procedure.
We use the $K$-means clustering algorithm in the second step.
We set the number of clusters to the number of true communities, run the $K$-means algorithm 10 times with different random seeds, and the select the best clustering in terms of the objective of the $K$-means algorithm (i.e., the mean squared distance between the nodes and their assigned cluster centroids).
Additionally, we also test an alternative data clustering method, Voronoi clustering, which assigns each node to the cluster with the closest centroid in the embedding space, the cluster centroids being the ones of the true communities.
Because the Voronoi clustering method has access to additional information about the locations of true communities, it provides the best-case scenario for the $K$-means algorithm.
The results for Voronoi clustering are presented in the Supporting Information 7.

We assess the performance by comparing the similarity between the planted partition of the network and the detected partition of the algorithm. We used the element-centric similarity~\cite{gatesElementcentricClusteringComparison2019}, denoted by $S$, with an adjustment such that a random shuffling of the community memberships for the two partitions yields $S=0$ on expectation (See Supporting Information Section 1). This way, for planted divisions into equal-sized communities, $S=0$ represents the baseline performance of the trivial algorithm, while $S>0$ indicates that communities are detectable by the given algorithm.

\subsection{Simulations: Planted Partition Model}

\begin{figure*}
    \centering
    \includegraphics[width=\textwidth]{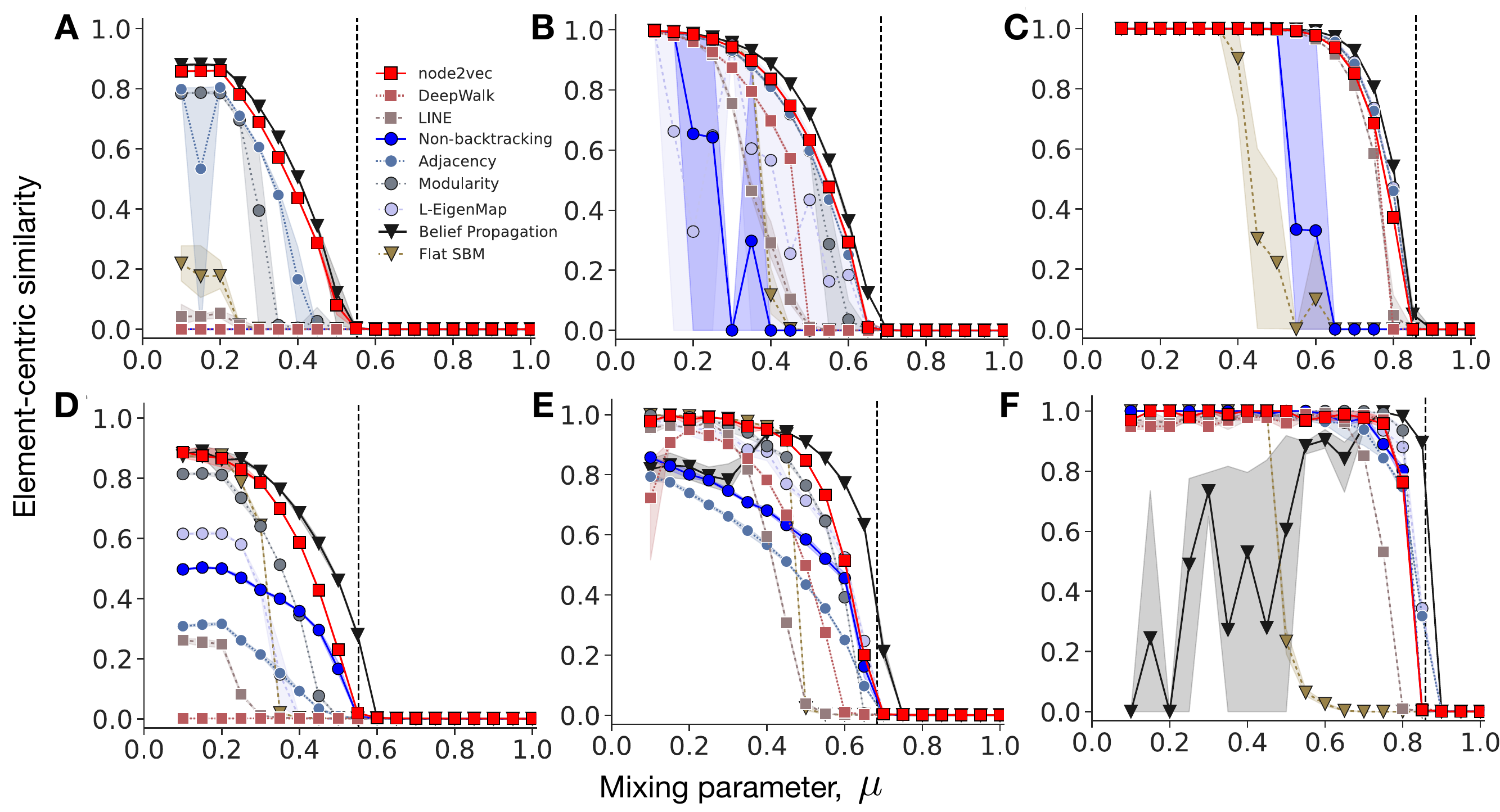}
    \caption{
        Performance of community detection methods for networks generated by the PPM as a function of the mixing parameter $\mu$.
        We generated networks with $n = 10^5$ nodes, different edge sparsity ($\langle k \rangle=5$ in A and D, $\langle k \rangle=10$ in B and E, $\langle k \rangle=50$ in C and F), and the different number of communities ($q=2$ for A--C and $q=50$ for D--F).
        The dashed vertical line indicates the theoretical detectability limit $\mu^*$ given by \eqref{eq:detectability_limit}:  communities are detectable (i.e., $S> 0$), in principle, below $\mu^*$.
        Spectral embedding methods detect communities up to the theoretical limit for dense networks (C and F), supporting the detectability limit derived by previous studies~\cite{nadakuditiGraphSpectraDetectability2012,RadicchiDetectabilityHeterogeneousNetworks2013}.
        However, for sparse networks, they fall short even at low $\mu$-values (A and D).
        node2vec and the spectral embedding based on the non-backtracking matrix outperform other spectral methods, with the performance curves close to that of the BP algorithm.
        Note that even the BP algorithm falls short of the exact recovery of some easily-detectable communities in the case of $q=50$ communities, with the initial parameters set according to the ground-truth communities.
        The error bands represent the 90\% confidence interval by a bootstrapping with $10^4$ resample.
        }
    \label{fig:performance_vs_mixing_multipartition}
\end{figure*}

We test the graph embedding and community detection algorithms on networks of $n=100,000$ nodes generated by the PPM, with $q \in \{2, 50\}$ communities of equal size and average degree $\langle k \rangle\in \{5, 10, 50\}$ (Fig.~\ref{fig:performance_vs_mixing_multipartition}).
Spectral methods find communities better than random guessing below the detectability limit $\mu^*$, i.e. $S > 0$, for $\mu < \mu^*$ and $\langle k \rangle=50$ (Figs.~\ref{fig:performance_vs_mixing_multipartition}C and F).
However, their performance is much worse when the average degree is small ($\langle k \rangle=5$, Figs.~\ref{fig:performance_vs_mixing_multipartition}A and D).
For example, Laplacian EigenMap falls short below the detectability limit ($\mu< \mu^*$), despite having the optimal detectability limit when the average degree is sufficiently large ~\cite{radicchiParadoxCommunityDetection2014}.
All techniques, including BP, which is supposed to be optimal for sparse networks, fail the exact recovery of the clusters for sparse networks even if the value of $\mu$ is low ($\langle k \rangle = 5$, Figs.~\ref{fig:performance_vs_mixing_multipartition}A and D).
We find that misclassifications are inevitable for these highly sparse networks because some nodes end up being connected with other communities more densely than with their own community by random chance.
The BP algorithm also fails for the networks with $q=50$ communities, even for small $\mu$ values.
The optimality of the belief propagation hinges on the assumption that loops in the network are negligible. This assumption is unlikely to be true when the network is dense, resulting in the underperformance of the BP algorithm.
Notably, the poor performance of the BP algorithm is mainly observed in the networks with 50 communities ($q=50$), where the prevalence of many local minima may exacerbate the limitations of the greedy optimization.

On the other hand, node2vec is substantially better than the spectral methods, and its performance is the closest to that of the BP algorithm for sparse networks (Figs.~\ref{fig:performance_vs_mixing_multipartition}A and D).
The results are striking given that the $K$-means algorithm can significantly worsen the performance of node2vec.
Crucially, the information theoretical limit of community detectability sharply separates the detectable and undetectable regime of communities for node2vec, demonstrating the validity of our theoretical result.
node2vec consistently achieves a good performance across different numbers of communities and different network sparsity.
Furthermore, node2vec performs well even if we reduce the embedding dimension $C$ from $64$ to $16$, which is smaller than the number of communities in the cases where $q=50$ (Supporting Information Section 5).
We also confirmed that the effectiveness of node2vec is robust for different sets of hyperparameter values (Supporting Information Section 6).


\subsection{Simulations: LFR benchmark}

\begin{figure*}
  \includegraphics[width=\hsize]{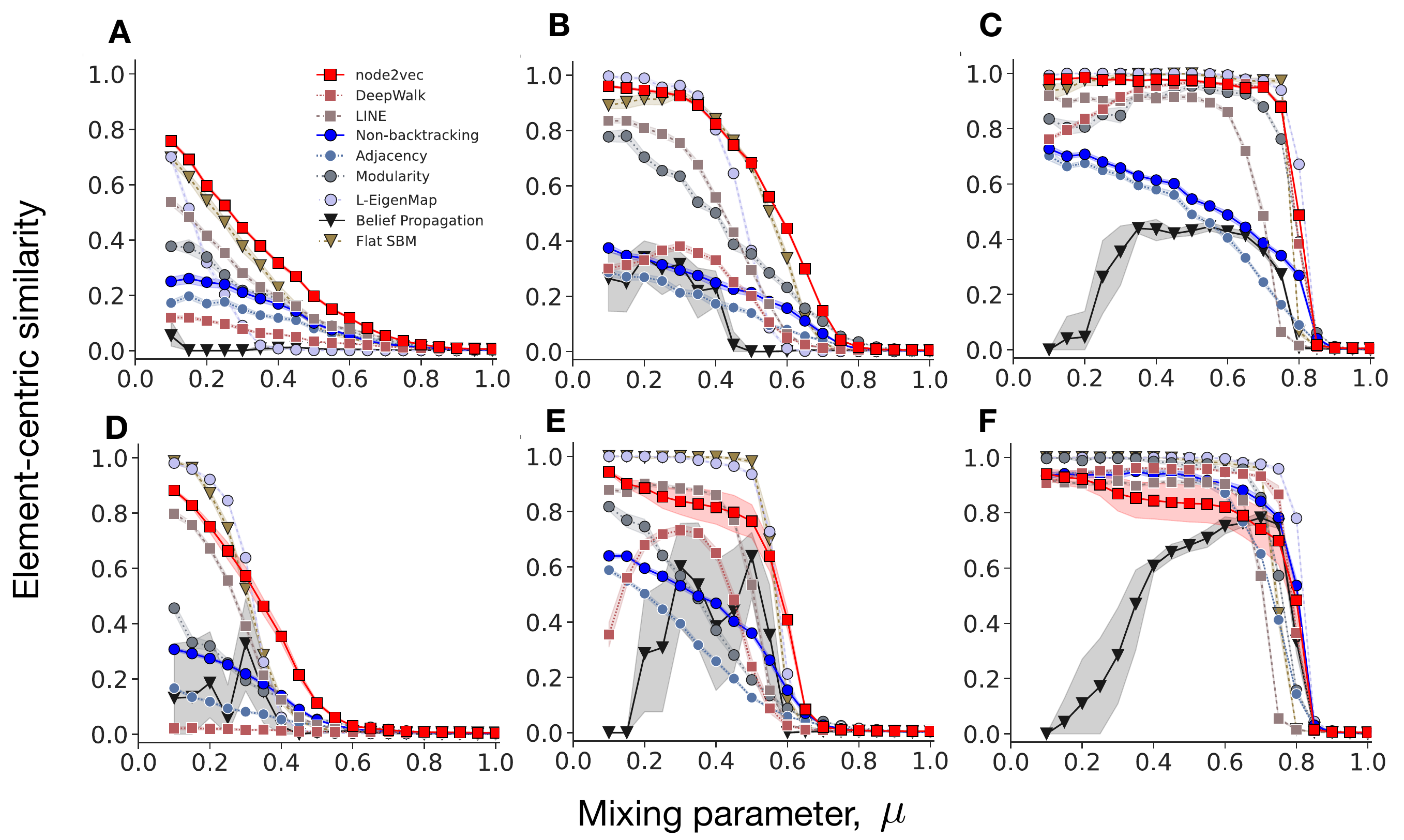}
    \caption{%
        Performance of community detection methods on the LFR benchmark networks, as a function of the mixing parameter $\mu$.
        We generated networks with $n = 10^4$ nodes with different edge sparsity ($\langle k \rangle=5$ in A and D, $\langle k \rangle=10$ in B and E, $\langle k \rangle=50$ in C and F).
        The degree exponent $\tau_1=2.1$ in A, B, and C, and $\tau_1=3$ in D, E, and F. node2vec consistently performs well across different sparsity regimes for most $\mu$-values, with a larger margin for sparser networks.
        The BP algorithm, which is provably optimal for networks generated by the PPM, fails to identify some easily-detectable communities, even with the initial parameters set according to the ground-truth communities.
        The error bands represent the 90\% confidence interval by a bootstrapping with $10^4$ resample.
    }
    \label{fig:lfr_performance_vs_mixing}
\end{figure*}

The PPM is a stylized model that lacks key characteristics of empirical community structure.
We test the graph embedding using more realistic networks generated by the LFR model~\cite{lancichinettiBenchmarkGraphsTesting2008}, which produces networks with heterogeneous degree and community size distributions, to assess the performance of the methods in a more practical context.
Unlike the PPM, however, the theoretical detectability limit of communities in LFR networks is not known.
We build the LFR networks by using the following parameter values: number of nodes $n=10,000$, degree exponent $\tau_1 \in \{2.1, 3\}$, average degree $\langle k \rangle\in \{5,10,50\}$, maximum degree $\sqrt{10n}$, community-size exponent $\tau_2 = 1$, community size range $[50, \sqrt{10n}]$.

In LFR networks, the BP algorithm and the non-backtracking embedding---which have an excellent performance on the PPM networks, at least in theory---underperform noticeably (Fig.~\ref{fig:lfr_performance_vs_mixing}), suggesting that
optimal methods for the standard PPM may not perform well in practice.
The underperformance is likely due to the violation of the assumption that loops are negligible in the network.
Even if the network is highly sparse, loops are likely to be formed
when the degree distribution is highly heterogeneous~\cite{Bianconi2005LoopsOA,Cantwell2023}.
As a result, the BP falls short for the LFR networks.
node2vec struggled to recover the planted communities perfectly, even when they were well-separated as previously noted~\cite{tandonCommunityDetectionNetworks2021}.
We note that the substandard performance of node2vec on the LFR networks may be attributed to the heterogeneity in community sizes, as the $K$-means algorithm tends to detect communities of nearly equal sizes~\cite{Wu2007AGO} (Figs.~\ref{fig:lfr_performance_vs_mixing}A and B).
In fact, when the Voronoi clustering method is used, the performance of node2vec is significantly improved, suggesting that
the substandard performance of node2vec is attributed to the clustering algorithm, not to the embedding itself.
Laplacian EigenMap outperformed other methods, except in extremely sparse networks (Figs.~\ref{fig:lfr_performance_vs_mixing}A and B).
It is worth noting that Laplacian EigenMap is highly sensitive to the number of dimensions. When the number of dimensions is set to $16$, Laplacian EigenMap underperforms considerably.
On the other hand, node2vec consistently performs well even across different number of dimensions (Supporting Information Section 7).
Even with the smaller embedding dimension $C=16$, node2vec performs comparably well with the flat SBM (Supporting Information Section 5).
We also confirmed that the effectiveness of node2vec is robust for different sets of hyperparameter values (Supporting Information 6).


\subsection{Empirical networks}

\begin{figure*}
    \centering
    \includegraphics[width=\textwidth]{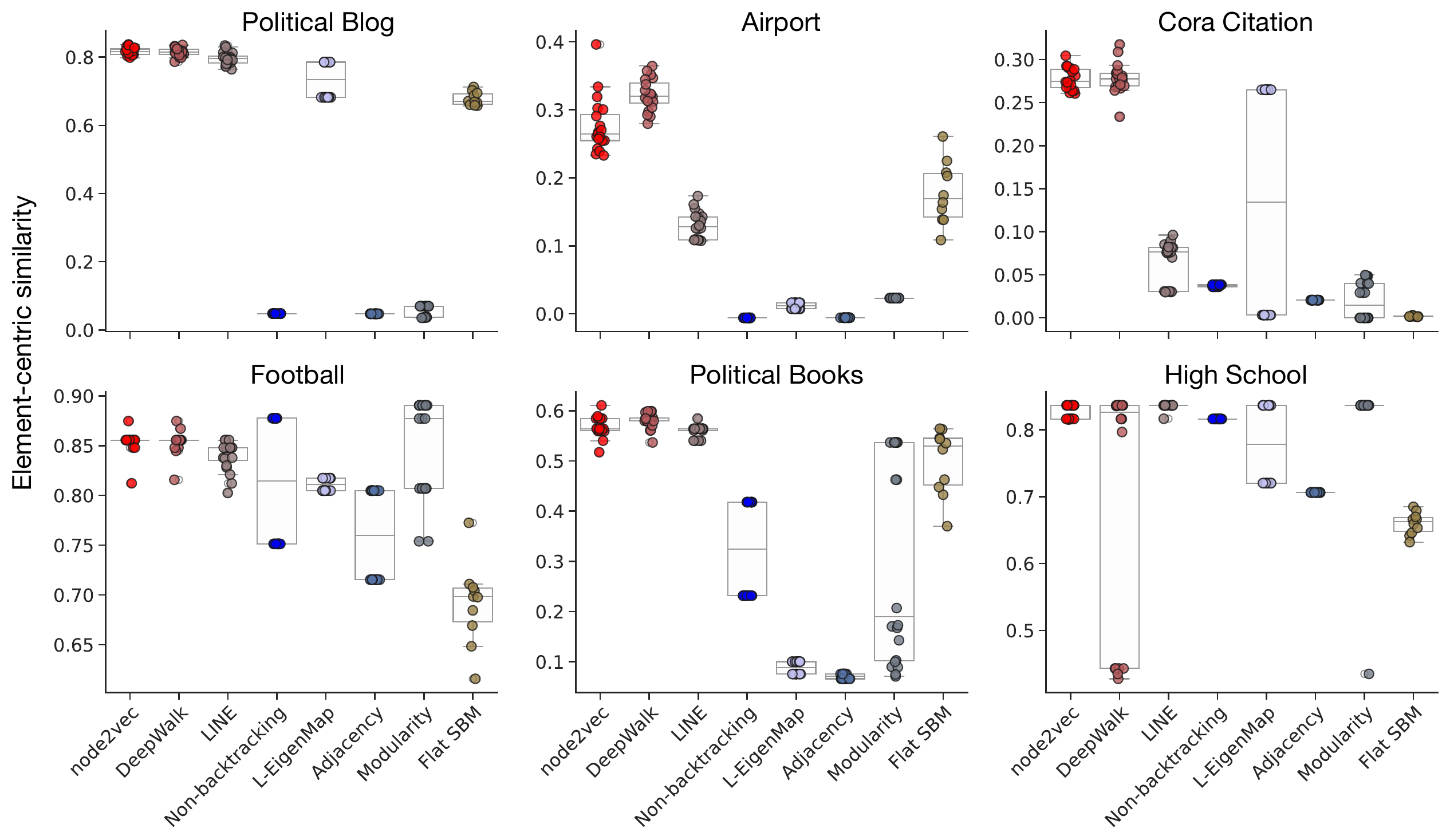}
    \caption{%
        Performance of community detection methods on empirical networks.
        Each panel illustrates the distribution of element-centric similarities for the community detection and graph embedding methods. Each circle denotes the outcome of a single run. The boxes indicate the quartiles of this distribution. The whiskers extend to the farthest data point within 1.5 times the interquartile range from the nearest hinge.
    }
    \label{fig:empirical}
\end{figure*}

We evaluated graph embedding methods using six empirical networks from various domains. Since communities in empirical networks are unknown, we relied on node metadata labels to define community memberships.
We note that node attributes do not necessarily align with the detected structural communities, hence community detection methods may fail to identify the node groups based on node attributes~\cite{Hric2014,peel2017ground}.
Keeping this potential issue in mind, we focus on the following networks, where node attributes align relatively well with the community structures, to shed light on the practical performance of graph embedding methods.
\texttt{Political blog network} represents hyperlinks between U.S. political blogs related to the 2004 U.S. presidential election~\cite{adamic2005political}.
The network consists of 1,222 nodes (blogs) and 16,714 edges, where an edge represents a citation from one blog to another on its front page.
As the community membership of the blogs, we use the blog categorization into liberal or conservative identified by an automated classification from several weblog directories.
\texttt{World-wide airport network} consists of 2,939 nodes representing airports in the world and 15,677 edges representing direct scheduled flight between the airports~\cite{Opsahl2011}.
As the community membership of the airports, we use the geographical classification into four regions (Africa, Americas, Asia \& Oceania, and Europe).
\texttt{Cora citation network} consists of 2,708 scientific publications and 5,429 citations among the publications~\cite{mccallum2000automating}.
As the community membership of the publications, we use the scientific field classification into seven fields of study (computer science, mathematics, physics, statistics, engineering, materials science, and medicine).
\texttt{Football network} represents American football games between Division IA colleges during regular season Fall 2000.
The nodes represent football teams and the edges represent the matches between the two teams. Each team belongs to one of 12 conferences, and we use the conference classification as the community membership~\cite{girvan2002community}.
\texttt{Political book network} represents a network of books on US politics published around the time of the 2004 presidential election. Each node represents a book, and two books are connected if they are frequently copurchased by the same buyers~\cite{newman2006modularity}. We use the political leaning of the books as the community membership.
\texttt{High-school network} represents a contact network of students in a high school in Marseilles, France. Each node represents a student and an edge between two students indicates a contact between them during 4 days in Dec. 2011~\cite{fournet2014contact}. The community membership of the students is the year of their high-school entrance.

We consider a scenario where the number $q$ of communities is not known.
We estimate $q$ by using the silouette score~\cite{rousseeuw1987silhouettes}.
Specifically, we identify the clusters with the $K$-means algorithm, by imposing a number of clusters $q$ going from $2$ to $20$, and choose the value of $q$ with the highest silouette score.
We use the same parameter set to generate the embeddings and identify the communities.
We run the whole process of community detection---from graph embedding, the estimation of the number of communities, and clustering---10 times with different random seeds, and report the agreement between the ground-truth communities and the detected communities in terms of the element-centric similarity for each run (Fig.~\ref{fig:empirical}).

node2vec and DeepWalk performed the best in four out of the six networks (\texttt{Political Blog}, \texttt{Airport}, \texttt{Cora Citation}, \texttt{Political Books}), and at least on par with the top-performing method in \texttt{High School}, suggesting that they performed consistently well across different networks (Fig.~\ref{fig:empirical}) .
Another neural embedding method---LINE---performed similary with node2vec and DeepWalk except for two networks (\texttt{Airport} and \texttt{Cora}).
The performance of the spectral embedding methods is less consistent across networks. For example, L-EigenMap can perform on par with the top-performing methods in three networks (\texttt{Political Blog}, \texttt{Cora}, and \texttt{High School}) but underperformed on the other four networks. Similary, Modularity embedding performed particularly well on \texttt{Football} but substantially underperformed on the other networks.

\section{Discussion}

We investigated the ability of neural graph embeddings to encode communities by focusing on shallow linear graph neural networks---node2vec, DeepWalk, and LINE---and comparing them with traditional spectral approaches.
We proved that, for not too sparse networks created by the PPM, node2vec is an optimal method to encode their community structure, in that the algorithmic detectability limit coincides with the information-theoretic limit.
Our results elucidate how and why node2vec works for community detection by demonstrating the equivalence between the embedding learned by node2vec and the spectral embedding based on the eigenvectors of the normalized Laplacian matrix. This equivalence provided insights into how communities in a network are embedded and the effectiveness of node2vec in learning network communities.

Our theoretical framework shows that graph embeddings based on simple neural networks can achieve optimal community detection. This finding provides guiding principles for developing effective neural embedding methods that are able to resolve communities in embedding space.
In neural graph embeddings, deep neural structures and non-linear activation are considered indispensable in order to achieve high performance.
The neural network architecture is also critical for graph neural networks for the community detection task~\cite{kawamotoMeanfieldTheoryGraph2018}.
Our findings instead demonstrate that a simple neural network with only one hidden layer and no non-linear activation can achieve the information-theoretical detectability limit of communities.

DeepWalk~\cite{perozziDeepWalkOnlineLearning2014} and LINE~\cite{tangLINELargescaleInformation2015} are also optimal in terms of the detectability limit of communities (Supplementary Information Section 2).
However, node2vec surpasses both DeepWalk and LINE
in numerical tests, owing to two key features.
First, node2vec learns degree-agnostic embeddings, which are highly robust against degree heterogeneity~\cite{kojakuResidual2VecDebiasingGraph2021}. By contrast, DeepWalk tends to learn node degree as the primary dimension in the embedding space~\cite{kojakuResidual2VecDebiasingGraph2021}. Consequently, degree heterogeneity introduces considerable noise to the community structure in the DeepWalk embedding.
Second, LINE is a specific instance of node2vec with window size $T=1$~\cite{qiuNetworkEmbeddingMatrix2018}, and thus learns the dyadic relationships between nodes.
As is the case for node2vec, LINE is resilient to degree heterogeneity, and performed closely to node2vec for some networks in our simulations.
However, it did not perform as well as node2vec, and this discrepancy may be attributed to LINE's emphasis on learning stochastic and noisy dyadic relationships, as opposed to the indirect relationships that node2vec captures.

Our results come with caveats.
First, our numerical results do not report the limiting performance of the embedding methods, but rather the lower bound of the performance limited by the $K$-means algorithm.
With graph embedding methods, the performance of community detection depends on both the quality of the embedding and the performance of the subsequent data clustering procedure.
Consequently, the performance of the graph embedding methods can be limited by the $K$-means algorithm. For instance, a previous study~\cite{tandonCommunityDetectionNetworks2021} using the $K$-means algorithm demonstrated that node2vec did not perform as well as standard community detection methods for the LFR networks even if its hyperparameters are fine-tuned.
Consistently with this result, the performance of node2vec was suboptimal for the LFR networks in our analysis.
However, we note that the LFR networks---that produce communities of different sizes---are challenging for the $K$-means algorithm---that tends to detect communities of nearly equal sizes~\cite{Wu2007AGO}.
In fact, communities in LFR networks are still well separated in the embedding of node2vec, as knowing the position of the centroids of the planted communities leads to a very good performance (Supplementary Information 7).
Nevertheless, the clustering step is a critical limitation when using graph embedding for community detection.
An extended $K$-means algorithm that can handle imbalanced cluster sizes could be a potential solution to this issue~\cite{LiangImbalancedKmeans}.
Our results reveal that communities are accurately represented in the embeddings, which might be sufficient for applications that can benefit from community structure but do not require the clustering step, such as link prediction~\cite{Ghasemian2019StackingMF} and node classification~\cite{kojakuResidual2VecDebiasingGraph2021,pengNeuralEmbeddingsScholarly2021}.

Second, in our analytical derivations, we assumed that the average degree is sufficiently large, as is the case for the corresponding analysis of spectral modularity maximization~\cite{nadakuditiGraphSpectraDetectability2012}.
Thus, the optimality may not hold if networks are substantially sparse.
However, our simulations suggest that node2vec is resilient to network sparsity compared with traditional spectral embedding methods.
Understanding the factor inducing such resilience is left to future work.

Third, while we restricted ourselves to the community detection task, graph embeddings have been used for other tasks, including link prediction, node classification, and anomaly detection.
Investigating the theoretical foundation behind the performance of neural embeddings in other tasks is a promising research direction.

We believe that our study will provide the foundation for future studies that uncover the inner workings of neural embedding methods and bridge the study of artificial neural networks to network science.

\section{Methods}

\subsection{node2vec as spectral embedding}
node2vec learns the structure of a given network based on random walks.
A random walk traverses a given network by following randomly chosen edges and generates the sequence of nodes $x^{(1)}, x^{(2)}, \ldots$.
The sequence is then fed into skip-gram word2vec~\cite{mikolovDistributedRepresentationsWords2013}, which learns how likely it is that a node $j$ appears in the surrounding of another node $i$ up to a certain time lag $T$ (i.e., window length)
through
the conditional probability
\begin{align}
    P(x^{(t + \tau)}=j \given x^{(t)}=i, 1\leq |\tau| \leq T) = \frac{1}{Z}\exp(\vec{u}_i ^\top \vec{v}_j) ,
\end{align}
where
$\vec{u}_i \in \mathbb{R}^{C \times 1}$, $\vec{v}_j \in \mathbb{R}^{C \times 1}$, and $Z$ is a normalization constant.
Each node $i$ is associated to two vectors:
vector $\vec{u}_i $ represents the embedding of node $i$; $\vec{v}_i$ represents node $i$ as a context of other nodes.
Because the normalization constant is computationally expensive, node2vec uses
a heuristic training algorithm, i.e., negative sampling~\cite{mikolovDistributedRepresentationsWords2013}.
When trained with negative sampling, skip-gram word2vec is equivalent to a spectral embedding that factorizes matrix $\mat{R}^{\text{n2v}}$ with elements~\cite{levyNeuralWordEmbedding2014,qiuNetworkEmbeddingMatrix2018}:
\begin{align}
    R^{\text{n2v}}_{ij} = \log  \frac{1}{T}\sum_{\tau=1}^T\left[\frac{P(x^{(t+\tau)} = j\given x^{(t)} = i)}{P(x^{(t)} = j)}\right], \label{eq:node2vec_matrix}
\end{align}
in the limit of $C\rightarrow n$ with $T$ greater than or equal to the network diameter, where $P(x^{(t)} = i)$ is the probability that the $t^\text{th}$ node in the given sequence is node $i$ (see Supporting Information Section 3 for the step-by-step derivation).
Note that the two embedding vectors $\vec{v}_i$ and $\vec{u}_i$ generated by node2vec are parallel to each other because $\mat{R}^{\text{n2v}}$ is symmetric~\cite{levyNeuralWordEmbedding2014,qiuNetworkEmbeddingMatrix2018,kojakuResidual2VecDebiasingGraph2021}.

Leveraging this equivalence, we take another step forward to connect the result with random matrix theory, deriving the detectability limit of these methods for community detection.
While previous studies demonstrated that node2vec factorizes $\mat{R}^{\text{n2v}}$, it remains unclear about its spectral properties, which is crucial to derive the detectability limit.
Deriving the spectrum of $\mat{R}^{\text{n2v}}$ in a closed form is challenging.
In fact, to identify the spectral density analytically, $\mat{R}^{\text{n2v}}$ needs to be described by a linear combination of the matrix before the element-wise logarithmic transformation (i.e., $\frac{1}{T}\sum_{\tau=1}^T\left[\frac{P(x^{(t+\tau)} = j\given x^{(t)} = i)}{P(x^{(t)} = j)}\right]$), which is not straightforward due to the nonlinearity and element-wise nature of the transformation.
Here, we derive the spectral properties of $\mat{R}^{\text{n2v}}$ by approximating the element-wise logarithm with a linear function based on the assumption that the window length $T$ is sufficiently large. To demonstrate our argument, let us describe $R_{ij}^{\text{n2v}}$ in the language of random walks.
Given that the network is undirected and unweighted, the probability $P(x^{(t)}=j)$ corresponds to the long-term probability of finding the random walker at node $j$.
The probability $P(x^{(t+\tau)} = j \given x^{(t)} = i)$ refers to the transition of a walker from node $i$ to node $j$ after $\tau$ steps.
In the limit $\tau \rightarrow \infty$, the walker reaches the stationary state, and $P(x^{(t+\tau)} = j \given x^{(t)} = i)$ approaches $P(x^{(t)}=j)$. Thus, in the regime of a sufficiently large $T$,
we take the Taylor expansion of $R^\text{n2v} _{ij}=\log\left(1 + \epsilon_{ij}\right)$ around $\epsilon_{ij} = \sum_{\tau=1}^T P(x^{(t+\tau)}=j \given x^{(t)} = i)/[TP(x^{(t)} = j)] -1$ and obtain
\begin{align}
    R^\text{n2v} _{ij} \simeq \hat R^\text{n2v} _{ij}
    &:=\frac{1}{T} \sum_{\tau = 1}^T \left[\frac{P(x^{(t+\tau)}=j \given x^{(t)} = i)}{P(x^{(t)} = j)}\right] -1.
\end{align}
In matrix form,
\begin{align}
    \label{eq:n2v_matrix}
    \mat{\hat R}^\text{n2v}&= \frac{2m}{T}\left[\sum_{\tau=1}^T \left(\mat{D}^{-1}\mat{A}\right)^{\tau} \right]\mat{D}^{-1} - \mat{1}_{n\times n},
\end{align}
where $\mat{A}$ is the adjacency matrix, $\mat{D}$ is a diagonal matrix whose diagonal element $D_{ii}$ is the degree $k_i$ of node $i$, $m$ is the number of edges in the network, and $\mat{1}_{n\times n}$ is the $n\times n$ all-one matrix.
We used $P(x^{(t)} = j) = k_j/2m$ and $\left(\mat{D}^{-1}\mat{A}\right)^{\tau} _{ij} = P(x^{(t+\tau)}=j \given x^{(t)} = i)$, derived from the fact that
$P(x^{(t)} = j)$ is proportional to degree in undirected networks; $\mat{D}^{-1}\mat{A}$ is the transition matrix, whose $\tau$th power represents the random walk transition probability after $\tau$ steps.

\begin{figure}
    \centering
    \includegraphics[width=0.7\hsize]{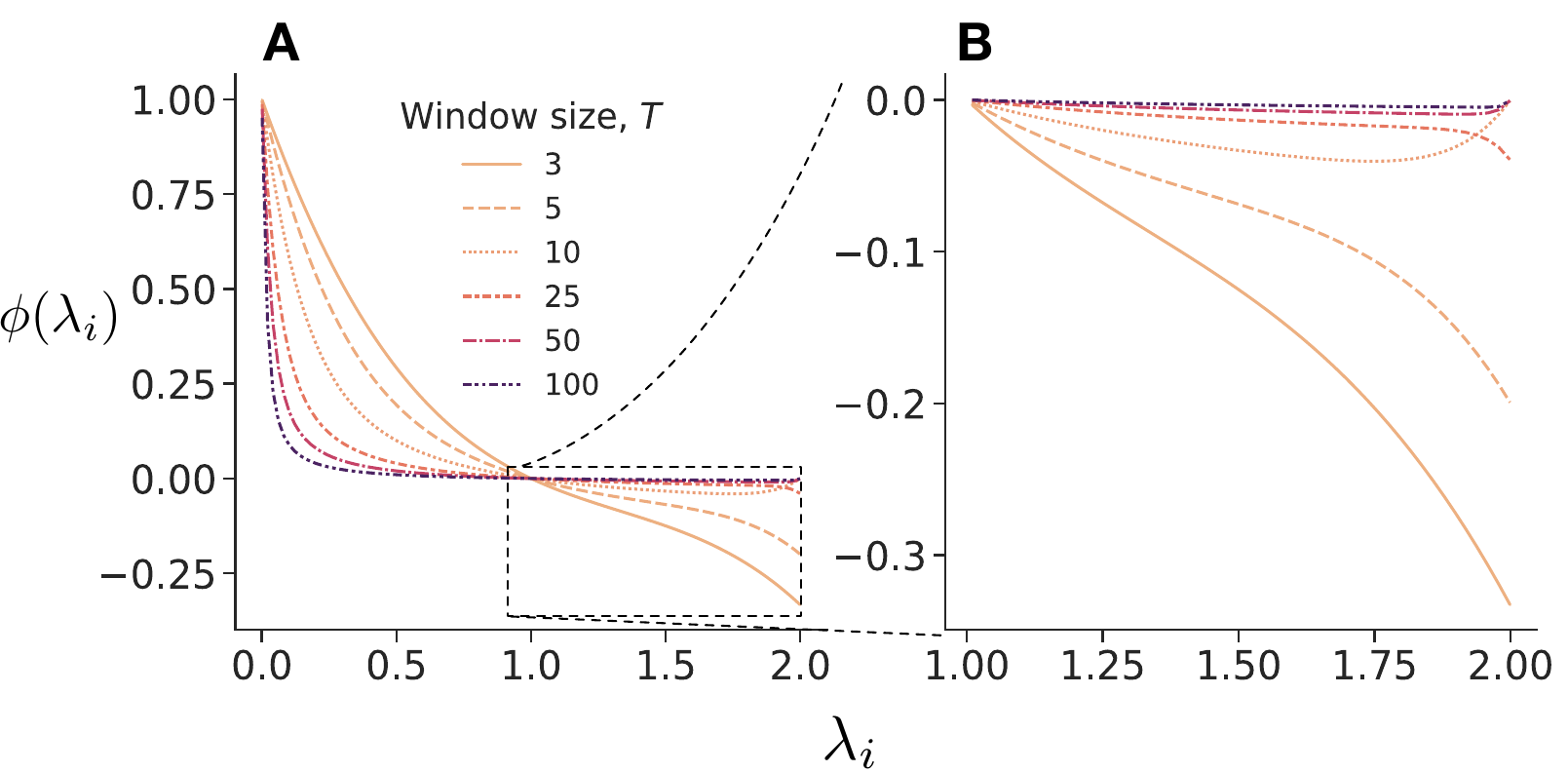}
    \caption{Graph kernel $\phi(\lambda_i; T)$ of node2vec matrix $\mat{\hat R}^{\text{n2v}}$ across different $T$ values.
    The function $\phi(\lambda_i)$ is non-negative and monotonically decreasing for $0 < \lambda_i \leq 1$ and $\phi(\lambda_i) \leq 0$ for $1 < \lambda_i \leq 2$.
    }
    \label{fig:phi}
\end{figure}

The node2vec matrix $\mat{\hat R}^\text{n2v}$ has a connection to the normalized Laplacian matrix, ${\mat{L}}$, which is tightly related to the characteristics of random walks and network communities~\cite{masudaRandomWalksDiffusion2017}.
The normalized Laplacian matrix is defined by ${\mat{L}}:=\mat{I} - \mat{D}^{-\frac{1}{2}}\mat{A}\mat{D}^{-\frac{1}{2}}$.
By using an alternative expression of the transition probability, i.e., $(\mat{D}^{-1}\mat{A})^\tau=\mat{D}^{-\frac{1}{2}}(\mat{D}^{-\frac{1}{2}}\mat{A}\mat{D}^{-\frac{1}{2}})^\tau\mat{D}^{\frac{1}{2}}$, we rewrite $\mat{\hat R}^{\text{n2v}}$ as
\begin{align}
    \mat{\hat R}^\text{n2v}&= \frac{2m}{T}\left[\sum_{\tau=1}^T \mat{D}^{-\frac{1}{2}}\left(\mat{D}^{-\frac{1}{2}}\mat{A}\mat{D}^{-\frac{1}{2}}\right)^{\tau}\mat{D}^{-\frac{1}{2}} \right] - \mat{1}_{n\times n} \nonumber \\
    &= 2m\mat{D}^{-\frac{1}{2}}\left[\frac{1}{T}\sum_{\tau=1}^T \left( \mat{I} - {\mat{L}} \right)^\tau
    - \frac{\mat{D}^{\frac{1}{2}}\vec{1}_{n}}{\sqrt{2m}}\frac{\mat{1}_{n} ^\top\mat{D}^{\frac{1}{2}}}{\sqrt{2m}}\right]\mat{D}^{-\frac{1}{2}},
    \label{eq:node2vec_matrix_approx}
\end{align}
where $\mat{1}_{n}$ is a column vector of length $n$.
We note that vector $\mat{D}^{1/2}\mat{1}_{n}/\sqrt{2m}$ is a trivial eigenvector of ${\mat{L}}$ associated with the null eigenvalue, $\lambda_1 = 0$.
Furthermore, $\left( \mat{I} - {\mat{L}} \right)^\tau$ changes the eigenvalues while keeping the eigenvectors intact.
This means that $\mat{\hat R}^\text{n2v}$ can be specified by using the spectrum of ${\mat{L}}$, i.e.,
\begin{align}
    \mat{\hat R}^{\text{n2v}}&=
    \mat{D}^{-\frac{1}{2}}\mat{\Gamma}
    \begin{bmatrix}
    \phi(\lambda_1) & & 0\\
     & \ddots & \\
    0 &  & \phi(\lambda_n) \\
    \end{bmatrix}
    \mat{\Gamma}^\top\mat{D}^{-\frac{1}{2}}, \label{eq:node2vec_matrix_form}
\end{align}
where $\mat{\Gamma} \in {\mathbb R}^{n\times n}$ is the matrix of the eigenvectors of ${\mat{L}}$, and $\phi$ is a \textit{graph kernel}~\cite{kunegisLearningSpectralGraph2009} that transforms the
eigenvalues $\lambda_i$ ($i=1,2,\ldots, n$) of ${\mat{L}}$ by
\begin{align}
\phi(\lambda_i)= \left\{
\begin{array}{ll}
\dfrac{2m(1-\lambda_i)\left[1-(1-\lambda_i)^{T}\right]}{T\lambda_i}  & (\lambda_i \neq 0),\\
0 & (\lambda_i=0),
\end{array} \right.
\label{eq:node2vec_graph_kernel}
\end{align}
or equivalently $\phi(\lambda_i) = \frac{2m}{T}\sum_{\tau = 1}^T (1-\lambda_i)^\tau$ if $\lambda_i\neq 0$ (Fig.~\ref{fig:phi}). Equation~(\ref{eq:node2vec_matrix_form}) tells us that the eigenvectors $\mat{U}$ of $\mat{\hat R}^{\text{n2v}}$ are equivalent to the eigenvectors $\mat{\Gamma}$ of the normalized Laplacian matrix, up to a linear transformation given by
\begin{align}
   \mat{U}:=\mat{D}^{\frac{1}{2}} \mat{\Gamma}.
\end{align}

Building on the correspondence between the normalized Laplacian ${\mat{L}}$ and the node2vec matrix $\mat{\hat R}^{\text{n2v}}$, we derive the algorithmic community detectability limit of node2vec.
Following \cite{nadakuditiGraphSpectraDetectability2012,radicchiParadoxCommunityDetection2014,RadicchiDetectabilityHeterogeneousNetworks2013},
 we assume that the network consists of two communities generated by the PPM.
Then, the non-trivial eigenvector of ${\mat{L}}$ encodes the communities and has the optimal detectability limit of communities, provided that the average degree is large (\eqref{eq:detectability_limit})~\cite{nadakuditiGraphSpectraDetectability2012,radicchiParadoxCommunityDetection2014,RadicchiDetectabilityHeterogeneousNetworks2013}.
This non-trivial eigenvector of $\mat{L}$ corresponds to the \textit{principal} eigenvector of $\mat{\hat R}^{\text{n2v}}$.
Specifically, the non-trivial eigenvector of $\mat{L}$ is associated with the smallest non-zero eigenvalue $\lambda_2$, which is $\lambda_2 < 1$ when each community is densely connected within itself and sparsely with other communities~\cite{vonluxburgTutorialSpectralClustering2007}.
The eigenvalues are mirrored in the eigenvalues $\phi(\lambda_i)$ of $\mat{\hat R}^{\text{n2v}}$, and $\lambda_2$---the smallest non-zero eigenvalue---yields the maximum $\phi$-value (Fig.~\ref{fig:phi})

This correspondence of non-trivial eigenvectors between $\mat{\hat R}^{\text{n2v}}$ and ${\mat{L}}$ suggests that communities detectable by ${\mat{L}}$ are also detectable by $\mat{\hat R}^{\text{n2v}}$ and vice versa.
Thus, spectral embedding with $\mat{\hat R}^{\text{n2v}}$ has the same information-theoretic detectability limit as spectral methods relying on eigenvectors of ${\mat{L}}$, for networks with sufficiently high degree.

\subsection{Detectability limit of DeepWalk}

We expand our argument to include DeepWalk~\cite{perozziDeepWalkOnlineLearning2014}.
Similar to node2vec, DeepWalk also trains word2vec but with a different objective function.
Previous studies have demonstrated that DeepWalk is a matrix factorization method~\cite{qiuNetworkEmbeddingMatrix2018,kojakuResidual2VecDebiasingGraph2021}. However, it remains unclear about the spectral properties of the matrix to be factorized. Furthermore, deriving the spectral properties of the matrix is challenging due to the element-wise logarithm involved in the matrix to be factorized.
More specifically, DeepWalk generates an embedding by factorizing a matrix with entries~\cite{kojakuResidual2VecDebiasingGraph2021}:
\begin{align}
R^\text{DW} _{ij} &:= \log \left(\frac{1}{T} \sum_{\tau=1}^T\frac{P(x^{(t)}=i, x^{(t + \tau)} = j)}{P(x^{(t)} = i)\cdot \frac{1}{n}} \right), \label{eq:deepwalk_R}
\end{align}
in the limit of $C\rightarrow n$ with $T$ being greater than the network diameter.
The element-wise logarithm in Eq.~\eqref{eq:deepwalk_R} makes it challenging to derive the spectral properties of $\mat{R}^\text{DW} _{ij}$.
Here, we employ a linear approximation by assuming that $T$ is sufficiently large.
When $T$ is large, the random walker reaches the stationary state, which is independent of where the walker starts from~\cite{masudaRandomWalksDiffusion2017}. Thus, we have
\begin{align}
\lim_{\tau \rightarrow \infty}P(x^{(t)}=i, x^{(t + \tau)} = j) &= P(x^{(t)}=i)P(x^{(t)}=j) \nonumber \\
&= P(x^{(t)}=i)\cdot \frac{k_j}{n \langle k \rangle} \label{eq:pij_deepwalk}
\end{align}
In particular, if the degree distribution is Poisson and the average degree is sufficiently large,
\begin{align}
\frac{k_j}{n \langle k \rangle} \simeq \frac{1}{n},\label{eq:stationary_prob_deepwalk}
\end{align}
which is true for the PPM.
By substituting \eqref{eq:stationary_prob_deepwalk} into \eqref{eq:pij_deepwalk}, we obtain
\begin{align}
P(x^{(t)}=i, x^{(t + \tau)} = j) \simeq P(x^{(t)}=i)\cdot \frac{1}{n} \quad \text{ for } \tau \gg 1.
\end{align}
Armed with this result, we demonstrate the detectability limit of DeepWalk as follows.
Assuming that the window length $T$ is large, we take the Taylor expansion of \eqref{eq:deepwalk_R}
around $\epsilon_{ij}' = \sum_{\tau=1}^T P(x^{(t)} = i, x^{(t+\tau)}=j)/[T(P(x^{(t)} = i) \cdot 1/n)] -1$ and obtain
\begin{align}
R^\text{DW} _{ij} \simeq {\hat R}^\text{DW} _{ij} &:= \left(\frac{1}{T} \sum_{\tau=1}^T\frac{P(x^{(t)}=i, x^{(t + \tau)} = j)}{P(x^{(t)} = i)\cdot \frac{1}{n}} \right) -1. \label{eq:deepwalk_R_approx}
\end{align}
In matrix form,
\begin{align}
\mat{\hat R}^{\text{DW}}:=\frac{n}{T}\left[\sum_{\tau=1}^T \left(\mat{D}^{-1}\mat{A}\right)^{\tau} \right] - \mat{1}_{n\times n}. \label{eq:deepwalk_R_approx_matrix}
\end{align}
Note that $\mat{\hat R}^{\text{DW}}$ is similar to the node2vec matrix $\mat{\hat R}^{\text{n2v}}$ (\eqref{eq:n2v_matrix}).
The right/left eigenvectors of $\mat{\hat R}^{\text{DW}}$ are obtained from those of the normalized Laplacian by simple multiplications by the operators $\mat{D}^{1/2}$ and $\mat{D}^{-1/2}$, respectively.
Therefore, DeepWalk has the information-theoretical detectability limit as well.

\subsection{Detectability limit of LINE}

LINE~\cite{tangLINELargescaleInformation2015} is a special version of node2vec with the window length being $T = 1$.
The corresponding matrix factorized by LINE is given by~\cite{qiuNetworkEmbeddingMatrix2018}:
\begin{align}
R^\text{LINE} _{ij} &:= \log \left( \frac{A_{ij}}{k_ik_j} + a_0\right) + \log 2m. \label{eq:line_R}
\end{align}
For LINE, although Ref.~\cite{qiuNetworkEmbeddingMatrix2018} shows $\log \left( \frac{A_{ij}}{k_ik_j}\right) + \log 2m$, we introduce a small positive value $a_0$  $(a_0> 0)$ to prevent the matrix elements from being infinite for $A_{ij}=0$.
To obtain the spectrum of $\mat{R}^\text{LINE}$,
we exploit the Taylor expansion $\log(x + a_0) \simeq \frac{x}{a_0} + \log a_0 $ around $x=0$, where $a_0>0$.
Specifically, assuming that the average degree is sufficiently large, we obtain
\begin{align}
\hat{R}^\text{LINE} _{ij} &= \frac{A_{ij}}{a_0 k_ik_j} + \log a_0 + \log 2m,
\end{align}
or equivalently in matrix form
\begin{align}
\mat{\hat R}^\text{LINE} &= \frac{1}{a_0}\mat{D}^{-1}\mat{A}\mat{D}^{-1} + a_1 \mat{1}_{n\times n} \nonumber \\
 &= \frac{1}{a_0}\mat{D}^{-1/2}(\mat{I} - \mat{L})\mat{D}^{-1/2} + a_1 \mat{1}_{n\times n} \nonumber \\
 &= \frac{1}{a_0}\mat{D}^{-1/2}\left(\mat{I} - \mat{L} + 2a_0a_1m \frac{\mat{D}^{\frac{1}{2}}\mat{1}_{n}}{\sqrt{2m}}\frac{\mat{1}_{n}^\top \mat{D}^{\frac{1}{2}}}{\sqrt{2m}}\right)\mat{D}^{-1/2}. \label{eq:line_matrix_approx}
\end{align}
where $a_1:= \log a_0 + \log (2m)$.
Equation~(\ref{eq:line_matrix_approx}) is reminiscent of \eqref{eq:node2vec_matrix_approx} for node2vec.
Comparing Eqs.~(\ref{eq:line_matrix_approx}) and (\ref{eq:node2vec_matrix_approx}), it immediately follows that
they share the same eigenvectors, and thus node2vec and LINE have the same detectability threshold.

\subsection*{Data availability}
The dataset used in this study is available in the Figshare database under accession code 10.6084/m9.figshare.26808775. The data can be obtained at ~\cite{KojakuDataset2023}.

\subsection*{Code availability}
We made available the code and documentations to reproduce all results. See our archived code at~\cite{github_archived} for reproducing our results and the up-to-date version at \cite{github} for replications.

\section*{Acknowledgements} \label{sec:ack}
This project was partially supported by the Army Research Office under contract number W911NF-21-1-0194, by the Air Force Office of Scientific Research under award numbers FA9550-19-1-0391 and FA9550-21-1-0446, and by the National Science Foundation under award numbers 1927418, and by the National Institutes of Health under awards U01 AG072177 and U19 AG074879.

\section*{Author Contributions Statement}
S.K. and F.R. performed the analysis and experiments.
S.K., F.R., Y.A., and S.F. conceived the research, discussed, and wrote the manuscript.

\section*{Competing Interest Statement}
The authors have no competing interest.


\end{document}



\title{Supporting Information: Network community detection via neural embeddings} 
\date{}
\author{Sadamori Kojaku, Filippo Radicchi, Yong-Yeol Ahn, Santo Fortunato}
\maketitle 

\tableofcontents

\section{Adjusted element-centric similarity}

We adjusted the original definition of the element-centric similarity in Ref.~\cite{gatesElementcentricClusteringComparison2019} such that the score for the two random partitions is zero.
In the following section, we define the element-centric similarity and its expected value for random partitions. Then, we define the adjusted element-centric similarity.

\subsection{Element-centric similarity for partitions}

Element-centric similarity (ECS) quantifies the difference between two partitions of nodes.
Let us represent a partition via the membership variables $\vec{g} = \{g_i\}_i$ $(i=1,2,\ldots,n)$, where $n$ is the number of nodes.
In the following three steps, ECS computes the similarity of two partitions $\vec{g}$ and $\vec{g}'$.
First, ECS constructs the \textit{affinity graph} for each partition. In the affinity graph for partition $\vec{g}$, two nodes $(i,j)$ are connected by an edge if they belong to the same community (i.e., $g_i = g_j$). Otherwise, $i$ and $j$ are not directly connected.
Second, ECS computes the neighborhood of each node by using a random walk. The random walk has a probability $\alpha$ of restarting the walk from the starting node. Because a node is connected to all other nodes in the same group in the affinity graph, the transition probability $p_{ij}^\vec{g}$ from $i$ to $j$ is given by
\begin{align}
p_{ij}^\vec{g} = \left\{
\begin{array}{ll}
\dfrac{\alpha}{n^\vec{g}_{g_i}}  + \delta_{ij}(1-\alpha) & (\text{if $i$ and $j$ belong to community $g$}) \\
0 & (\text{otherwise}),
\end{array}
\right. \label{eq:paij}
\end{align}
where $n^{\vec{g}} _{g_i}$ is the number of nodes in group $g_i$ in partition $\vec{g}$, and $\delta_{ij}$  is Kronecker delta.
Third, SCE deems two partitions $\vec{g}$ and $\vec{g}'$ as similar if the respective transition probabilities $p_{ij}^\vec{g}$ and $p_{ij} ^{\vec{g}'}$ are similar, i.e.,
\begin{align}
S(\vec{g}, \vec{g}') := 1 - \frac{1}{2n\alpha}\sum_{i=1}^n \sum_{j=1}^n |p^{\vec{g}}_{ij}-p^{\vec{g}'}_{ij}|.
\label{eq:ECSstd}
\end{align}
By substituting Eq.~\eqref{eq:paij} into Eq.~\eqref{eq:ECSstd}, we obtain
\begin{align}
S(\vec{g}, \vec{g}')
&= 1 - \frac{1}{2n\alpha}\sum_{i=1}^n\sum_{j=1}^n |p^{\vec{g}}_{ij}-p^{\vec{g}'}_{ij}| = 1 - \frac{1}{2n}\sum_{i=1}^n\sum_{j=1}^n \left|\frac{\delta_{g_i,g_j}}{n^\vec{g} _{g_i}}-\frac{\delta_{g_i',g_j'}}{n^{\vec{g}'} _{g_i'}}\right|\nonumber \\
&= 1 - \frac{1}{2n}
\sum_{i=1}^n \sum_{j=1}^n \left[\frac{\delta_{g_i, g_j}}{n^{\vec{g}}_{g_i}}
+ \frac{\delta_{g_i ' g_j' }}{n^{\vec{g}'}_{g'_i}}+ \delta_{g_i g_j}\delta_{g_i ' g_j '}\left(
\left| \frac{1}{n^{\vec{g}}_{g_i}} - \frac{1}{n^{\vec{g}'}_{g'_i}} \right|
 - \frac{1}{n^{\vec{g}}_{g_i}} - \frac{1}{n^{\vec{g}'}_{g'_i}} \right)
\right]. \nonumber \\
&= 1 - \frac{1}{2n}\left[
n+n
+ \sum_{i=1}^n \sum_{j=1}^n\delta_{g_i g_j}\delta_{g_i ' g_j '}
\left(
\left| \frac{1}{n^{\vec{g}}_{g_i}} - \frac{1}{n^{\vec{g}'}_{g'_i}} \right|
 - \frac{1}{n^{\vec{g}}_{g_i}} - \frac{1}{n^{\vec{g}'}_{g'_i}} \right)
\right] \nonumber \\
&=\frac{1}{2n}\sum_{i=1}^n \sum_{j=1}^n\delta_{g_i g_j}\delta_{g_i ' g_j '}\left(
  \frac{1}{n^{\vec{g}}_{g_i}} + \frac{1}{n^{\vec{g}'}_{g'_i}} - \left| \frac{1}{n^{\vec{g}}_{g_i}} - \frac{1}{n^{\vec{g}'}_{g'_i}} \right|
\right)  \nonumber \\
&=
\dfrac{1}{2n}
\sum_{i=1}^n \sum_{j=1}^n
\delta_{g_i g_j}\delta_{g_i ' g_j '}
\left\{
\begin{array}{ll}
    \dfrac{2}{n_{g'_i} ^{\vec{g}'}}
   &\left(\frac{1}{n^{\vec{g}}_{g_i}} > \frac{1}{n^{\vec{g}'}_{g'_i}}\right) \nonumber \\
    \dfrac{2}{n_{g_i} ^{\vec{g}}}
   &\left(\frac{1}{n^{\vec{g}}_{g_i}} \leq \frac{1}{n^{\vec{g}'}_{g'_i}}\right)
\end{array}
\right. \nonumber \\
&=
\dfrac{1}{n}
\sum_{i=1}^n \sum_{j=1}^n
\delta_{g_i g_j}\delta_{g_i ' g_j '} \min\left(\frac{1}{n_{g_i}^{\vec{g}}}, \frac{1}{n_{g_i '}^{\vec{g}'}}\right) \nonumber \\
&=\frac{1}{n}\sum_{c=1}^{C^{\vec{g}}}\sum_{c'=1}^{C^{\vec{g}'}}  \left(n^{\vec{g}, \vec{g}'} _{c, c'}\right)^2\min \left(
 \frac{1}{n^{\vec{g}}_c},  \frac{1}{n^{\vec{g}'}_{c'}}
\right), \label{eq:ecs}
\end{align}
where $n^{\vec{g}, \vec{g}'} _{c,c'}$ is the number of nodes that belong to group $c$ in partition $\vec{g}$ and group $c'$ in partition $\vec{g}'$, and $C^{\vec{g}}$ and $C^{\vec{g}'}$ are the number of groups in partition $\vec{g}$ and $\vec{g}'$, respectively.
We note that the restarting probability $\alpha$ is canceled and does not affect the similarity.

\subsection{Element-centric similarity for random partitions}
We derived the element-centric similarity between a given partition $\vec{g}$ and random partitions $\vec{\zeta}$.
We generate the random partition by shuffling the group membership. This randomization preserves the number of groups and the size of each group. In the random partition, a node belongs to a group $c'$ of size $n^{\vec{\zeta}}_{c'}$ with probability $n^{\vec{\zeta}} _{c'}/n$. Thus, the expected number of nodes in group $c$ in partition $\vec{g}$ that belong to group $c'$ in the random partition is given by
\begin{align}
\Exp_{\vec{\zeta}}\left[n_{c, c'}^{\vec{g}, \vec{\zeta}}\right] = \sum_{i=1}^n \delta_{g_i, c}\frac{n_{c'} ^{\vec{g}}}{n} = \frac{n_c ^{\vec{g}}n_{c'} ^{\vec{g}}}{n}. \label{eq:nccprime}
\end{align}
By substituting Eq.~\eqref{eq:nccprime} into Eq.~\eqref{eq:ecs}, we have
\begin{align}
{\mathbb E}_{\vec{\zeta}}[S(\vec{g}, \vec{\zeta})]&=
\frac{1}{n}
\sum_{c=1}^{C^{\vec{g}}} \sum_{c'=1}^{C^{\vec{g}}} \left(
\frac{n_c ^{\vec{g}}n_{c'} ^{\vec{g}}}{n}\right)^2
\min \left(
 \frac{1}{n^{\vec{g}}_c},  \frac{1}{n^{\vec{g}}_{c'}}
\right) \nonumber \\
&=\frac{1}{n^3}\sum_{c=1}^{C^{\vec{g}}} \sum_{c'=1}^{C^{\vec{g}}} n_c ^{\vec{g}}n_{c'} ^{\vec{g}}
\min \left(n^{\vec{g}}_c,  n^{\vec{g}}_{c'}\right) \nonumber \\
&=\sum_{c=1}^{C^{\vec{g}}} \sum_{c'=1}^{C^{\vec{g}}} z_c ^{\vec{g}}z_{c'} ^{\vec{g}}
\min \left(z^{\vec{g}}_c,  z^{\vec{g}}_{c'}\right),
\end{align}
where $z^{\vec{g}}_c = n^{\vec{g}}_c/n$ is the fraction of nodes in group $c$ in partition $\vec{g}$.

\subsection{Normalized element-centric similarity}
We adjusted the element-centric similarity such that random partitions have a score of zero, i.e.,
\begin{align}
    \overline S(\vec{g}, \vec{\zeta}) = \frac{ S(\vec{g}, \vec{g}') - \Exp_{\vec{\zeta}}\left[ S(\vec{g}, \vec{\zeta})\right]}{1 -  \Exp_{\vec{\zeta}}\left[ S(\vec{g}, \vec{\zeta})\right]}.
\end{align}

\section{Reparameterization of detectability limit}
\label{sec:detectability_limit_reparameterization}

In Ref.~\cite{nadakuditiGraphSpectraDetectability2012}, the detectability limit for the spectral embedding with $\mat{A}$ is described
using $c_\text{in}=np_\text{in}$ and $c_\text{out}=np_\text{out}$ as
\begin{align}
c_{\text{in}} - c_{\text{out}} > \sqrt{nq(c_\text{in} + (q-1)c_\text{out})},
\end{align}
First, we rewrite the inequality using $\langle k \rangle$, $p_{\text{in}}$, $p_{\text{out}}$ as
\begin{align}
&c_{\text{in}} - c_{\text{out}} > \sqrt{nq(c_\text{in} + (q-1)c_\text{out})} \\
&\Rightarrow
np_{\text{in}} - np_{\text{out}} > q\sqrt{\frac{1}{q}np_\text{in} + \left(1-\frac{1}{q}\right)np_\text{out}} \nonumber \\
&\Rightarrow
np_{\text{in}} - np_{\text{out}} > q\sqrt{\langle k \rangle}, \label{eq:kp2}
\end{align}
where we have exploited
\begin{align}
\langle k \rangle = q^{-1}np_\text{in} + (1-q^{-1})np_\text{out}. \label{eq:kp}
\end{align}
By rearranging Eq.~\eqref{eq:kp} into $np_{\text{in}} = q \langle k \rangle - (q-1) np_{\text{out}}$ and substituting it into Eq.~\eqref{eq:kp2}, we obtain
\begin{align}
&np_{\text{in}} - np_{\text{out}} > q\sqrt{\langle k \rangle} \nonumber \\
&\Rightarrow
\left[q \langle k \rangle - (q-1) np_{\text{out}}\right]- np_{\text{out}} > q\sqrt{\langle k \rangle} \nonumber \\
&\Rightarrow
q (\langle k \rangle - np_{\text{out}}) > q\sqrt{\langle k \rangle} \nonumber \\
&\Rightarrow
\langle k \rangle - \mu \langle k \rangle > \sqrt{\langle k \rangle} \nonumber \\
&\Rightarrow
\mu < \mu^*,\; \text{where}\;\; \mu^* := 1 - \frac{1}{\sqrt{\langle k \rangle}},
\end{align}
where we remind that $\mu = np_{\text{out}} /\langle k \rangle$.

\section{node2vec as matrix factorization}
\label{sec:spectral_node2vec}

node2vec embedding is generated from a sequence of nodes $\{x^{(1)}, x^{(2)}, \ldots \}$ obtained from random walks in a network. We assume that the sequence is infinite in its length.
A window of length $2T + 1$ slides over the sequence, with the center node at the $T^{\text{th}}$ position and $T$ nodes before and after it, i.e.,
\begin{align}
 \ldots, x^{(t-T-1)}, \overbrace{\underbrace{x^{(t-T)}, \ldots, x^{(t-1)}}_{\text{context}}, \underbrace{x^{(t)}}_{\text{center}}, \underbrace{x^{(t+1)}, \ldots, x^{(t + T)}}_{\text{context}}}^{\text{Sliding window}}, x^{(t+T+1)}, \ldots
\end{align}
This sequence then trains the skip-gram word2vec using negative sampling~\cite{mikolovDistributedRepresentationsWords2013}.
Negative sampling learns a correlational association between the center and context nodes in light of a random correlation.
More specifically, consider the conditional probability $P(x^{(t+\tau)}=j \vert x^{(t)}=i)$ that node $j$ appears after $\tau$ steps from the center node $x^{(t)} = i$.
This probability is strongly correlated with the frequency of $j$ in the entire sequence---$P(x^{(t+\tau)}=j)$---because a frequent node in the given sequence also frequently appears in the window.
Negative sampling discounts this frequency effect by contrasting the context $j$ with
a random node $j'$ sampled from the given sequence.

Operationally, one generates a list ${\cal D}$ of node pairs to train the word2vec model.
List ${\cal D}$ is a union of two lists ${\cal D}^{\text{data}}$ and ${\cal D}^{\text{rand}}$.
${\cal D}^{\text{data}}$ includes the node pairs $(i,j)$ consisting of a center node $i$ and a context node $j$ that co-appear in the same window in the given sequence.
Another list ${\cal D}^{\text{rand}}$ includes the node pairs $(i,j')$ consisting of a center node $i$ sampled from the given sequence and a random node $j'$ sampled from a random distribution $P_0(j')$.
We use a typical random distribution, i.e., we use the long-term probability $P(x^{(t)} = j')$ of random walks as $P_0(j)$~\cite{qiuNetworkEmbeddingMatrix2018}.
Then, the skip-gram word2vec model estimates the probability $P\left((i,j) \in {\cal D}^{\text{data}}\right)$ that a given pair $(i,j)$ comes from ${\cal D}^{\text{data}}$ by
\begin{align}
    \sigma(\vec{u}_i ^\top \vec{v}_j) = \frac{1}{1 + \exp(-\vec{u}_i ^\top \vec{v}_j)},
\end{align}
where $\vec{u}_i$ and $\vec{v}_j$ are the column embedding vectors of center node $i$ and context node $j$, respectively.
The embedding vectors are determined by maximizing the log-likelihood
\begin{align}
    {\cal J} &= \sum_{(i,j) \in {\cal D}^{\text{data}}} \log \sigma(\vec{u}_i^\top \vec{v}_j) +\sum_{(i,j) \in {\cal D}^{\text{rand}}} \log \left(1 - \sigma(\vec{u}_i^\top \vec{v}_j)\right).
\end{align}

The maximization of ${\cal J}$ can be translated into a matrix factorization problem~\cite{levyNeuralWordEmbedding2014}.
One parametrizes dot similarity $\vec{u}_i^\top \vec{v}_j$ as a single variable $R_{ij}$ and assumes that the elements $R_{ij}$ are independent of each other.
This assumption holds if the embedding dimension is sufficiently large~\cite{levyNeuralWordEmbedding2014,kojakuResidual2VecDebiasingGraph2021}.
By taking the derivative
\begin{align}
&\frac{\partial {\cal L}}{\partial R_{ij}} = \left| {\cal D}^{\text{data}} \right| P\left( (i,j) \in {\cal D}^{\text{data}}\right) (1-\sigma(R_{ij})) \nonumber \\
    &\quad \quad \quad - \left|{\cal D}^{\text{rand}}\right| P\left( (i,j) \in {\cal D}^{\text{rand}}\right) \sigma(R_{ij}),
\end{align}
and solving $\partial {\cal L}/ \partial R_{ij} = 0$, we obtain
\begin{align}
    \sigma(R_{ij}) = \frac{P\left( (i,j) \in {\cal D}^{\text{data}} \right)}{P\left( (i,j) \in {\cal D}^{\text{data}} \right) + \dfrac{\left| {\cal D}^{\text{rand}} \right|}{\left| {\cal D}^{\text{data}} \right|}P\left( (i,j) \in {\cal D}^{\text{rand}} \right)}.
\end{align}
We assume that $P\left( (i,j) \in {\cal D}^{\text{data}}\right), P\left( (i,j) \in {\cal D}^{\text{rand}}\right)>0$, which is true when the window size is larger than or equal to the diameter of the network.
Rearranging the equation yields
\begin{align}
    R_{ij} = \log \dfrac{P\left( (i,j) \in {\cal D}^{\text{data}} \right)}{  P\left( (i,j) \in {\cal D}^{\text{rand}} \right)} - \log \left| {\cal D}^{\text{rand}} \right|+ \log \left| {\cal D}^{\text{data}} \right|. \label{eq:Rij}
\end{align}
Now, let us specify $P\left( (i,j) \in {\cal D}^{\text{data}} \right)$ and $P\left( (i,j) \in {\cal D}^{\text{rand}} \right)$.
Remind that ${\cal D}^{\text{data}}$ is the node pairs
sampled from a random-walk sequence generated from the given network.
More specifically,
\begin{align}
P\left( (i,j) \in {\cal D}^{\text{data}} \right) = \sum_{\tau=-T, \tau \neq 0}^T P(x^{(t)}=i, x^{(t+\tau)}=j)P_\text{NS}(\tau),
\end{align}
where $P_{\text{NS}}(\tau)$ is the probability that the $\tau^{\text{th}}$ node is sampled as a context node and paired with the center node $i$.
Because each context node is sampled with the same probability, $P_\text{NS}(\tau)=1/2T$, which gives
\begin{align}
P\left( (i,j) \in {\cal D}^{\text{data}} \right) = \frac{1}{2T}\sum_{\tau=-T, \tau \neq 0}^T P(x^{(t)}=i, x^{(t+\tau)}=j). \label{eq:data}
\end{align}
Another list ${\cal D}^{\text{rand}}$ is created by two independent sampling processes, one process sampling a node $i$ from the given sequence and the other process sampling another node $j$ from the random distribution $P_0(j)$. The former process is essentially the same as the latter because $P_0(j)$ is proportional to the frequency of node $j$.
Thus, we have
\begin{align}
P\left( (i,j) \in {\cal D}^{\text{rand}} \right) = P_0(i)P_0(j),
\end{align}
or equivalently,
\begin{align}
P\left( (i,j) \in {\cal D}^{\text{rand}} \right) =P(x^{(t)}=i)P(x^{(t)}=j). \label{eq:prand}
\end{align}
Altogether, by substituting Eqs.~\eqref{eq:data} and \eqref{eq:prand} into Eq.~\eqref{eq:Rij}, we have
\begin{align}
R_{ij} = \log \left(\frac{1}{2T} \sum_{\tau = -T; \tau \neq 0}^T\frac{P(x^{(t)}=i, x^{(t + \tau)} = j)}{P(x^{(t)} = i)P(x^{(t)} = j)} \right) -\log \left| {\cal D}^{\text{rand}} \right|+ \log \left| {\cal D}^{\text{data}} \right|.
\end{align}
We can neglect the constant $-\log \left| {\cal D}^{\text{rand}} \right|+ \log \left| {\cal D}^{\text{data}} \right|$ because it does not change the non-trivial eigenvectors of $\mat{R}$. Thus, we obtain
\begin{align}
R_{ij} = \log \left(\frac{1}{2T} \sum_{\tau = -T; \tau \neq 0}^T\frac{P(x^{(t)}=i, x^{(t + \tau)} = j)}{P(x^{(t)} = i)P(x^{(t)} = j)} \right).
\end{align}
Random walks in undirected networks are reversible, i.e., $P(x^{(t)}=i, x^{(t + \tau)} = j) = P(x^{(t)}=i, x^{(t - \tau)} = j)$. Thus, we have
\begin{align}
R_{ij} = \log \left(\frac{1}{T} \sum_{\tau = 1}^T\frac{P(x^{(t + \tau)} = j \vert x^{(t)}=i)}{P(x^{(t)} = j)} \right),   \label{eq:node2vec_R}
\end{align}
in the main text.
By substituting $R_{ij}=\vec{u}_i^\top \vec{v}_j$ we obtain a matrix decomposition problem
\begin{align}
    \mat{R} = \mat{U} \mat{V}^\top,
\end{align}
where $\mat{R}\in {\mathbb R}^{n \times n}$ with element $R_{ij}$, $\mat{U} = [\vec{u}_1, \ldots,\vec{u}_n]^\top$, and $\mat{V} = [\vec{v}_1, \ldots,\vec{v}_n]^\top$.
Because $\mat{R}$ is a symmetric matrix, we can find such a decomposition by the eigendecomposition, i.e.,
\begin{align}
    \mat{U} = \mat{V} = \mat{\Lambda}^{1/2} {\bf\Gamma}, \label{eq:matrix_factorization_node2vec}
\end{align}
where ${\bf\Gamma}$ is the matrix of eigenvectors, and $\mat{\Lambda}$ is a diagonal matrix with the corresponding eigenvalues in the diagonals.
Equation~(\ref{eq:matrix_factorization_node2vec}) coincides with the optimal solution for the original objective ${\cal J}$ of word2vec, provided
that the embedding dimension $C$ is equal to the number of nodes. Even with a smaller $C$, Eq.~\eqref{eq:matrix_factorization_node2vec} provides a good approximation~\cite{qiuNetworkEmbeddingMatrix2018,kojakuResidual2VecDebiasingGraph2021}.

\section{Implementation}
\subsection{node2vec, LINE, and DeepWalk}
For node2vec, we set the length of a single walk to 80, the number of walkers per node to 40, the length of the window to 10, and the number of epochs to train to 1 while not biasing the random walk ($p=q=1$).
We show the performance of node2vec for a broad range of hyperparameter values in Supporting Information Section \ref{sec:robustness}.
We use the word2vec implemented in the \texttt{gensim} package~\cite{rehurek2011gensim} with the default parameters of version 4.3.
For LINE, we increase the number of walks to 400 because LINE is trained with fewer iterations than node2vec. Similarly, we increase the number of walks for DeepWalk to 120. We set the other parameters to those used in node2vec.

\subsection{Laplacian EigenMap}

We used the standard eigenvector solver---\texttt{scipy.linalg.eigs}---implemented in scipy~\cite{2020SciPy-NMeth} to compute the eigenvectors of the normalized Laplacian matrix.
However, the eigenvector solver did not converge for some networks due to numerical stability since some eigenvalues are very close to zero.
To facilitate the convergence, we shift the smallest eigenvalues to the largest values (which preserves the eigenvectors) via the following transformation~\cite{Jiang2023relations}:
\begin{align}
\mat{L}:= \mat{I} - \mat{D}^{-\frac{1}{2}} \mat{A}\mat{D}^{-\frac{1}{2}},
\end{align}
where $\mat{L}$ is the normalized Laplacian matrix, $\mat{I}$ is the identity matrix, $\mat{D}$ is a diagonal matrix whose $i$th diagonal entry $D_{ii}$ is the degree of node $i$, and $\mat{A}$ is the adjacency matrix.

Laplacian EigenMap relies on the eigenvectors of $\mat{L}$ with the non-zero smallest eigenvalues. Some of these smallest eigenvalues can be nearly zero, which is the cause of the numerical instability. Thus, we shift the eigenvalues by:
\begin{align}
\mat{\overline L} := 2\mat{I} - \mat{L} = \mat{I} + \mat{D}^{-\frac{1}{2}} \mat{A}\mat{D}^{-\frac{1}{2}}.
\end{align}
This transformation changes the eigenvalue $\lambda_i$ of $\mat{L}$ to $2 - \lambda_i$, while the eigenvectors remain unchanged. Because the eigenvalues of $\mat{L}$ are bounded in the range $[0,2]$, the smallest eigenvalues of $\mat{L}$ correspond to the largest eigenvalues of $\mat{\overline L}$, with a sufficient distance from zero. Thus, Laplacian EigenMap can be obtained by computing the eigenvectors associated with the largest eigenvalues of the shifted Laplacian matrix $\mat{\overline L}$.

\subsection{Modularity embedding}
\label{sec:mudularity_embedding}
The Modularity embedding is based on the eigenvectors of the modularity matrix, which is defined as
\begin{align}
\mat{Q}:= \frac{1}{2m}\left(\mat{A} - \frac{\vec{k}\vec{k}^\top}{2m} \right),
\end{align}
where $\vec{k}$ is a column vector of length $n$ with each element $k_i$ representing the degree of node $i$, and $m$ is the total number of edges in the network.
The modularity matrix $\mat{Q}$ is a fully dense matrix, and constructing this matrix for large networks requires a significant amount of memory. This makes it computationally challenging to construct the matrix and compute its eigenvectors for large networks.

To mitigate the computational challenge, we utilized a power-iteration method~\cite{newmanFindingCommunityStructure2006} that leverages the structure of the modularity matrix and scales with the network size.
The power iteration method finds the eigenvectors of the modularity matrix by iteratively applying the following transformation to a random vector $\vec{x}$ until convergence:
\begin{align}
    \vec{x} &\leftarrow \left(\mat{Q} - \mat{S}\right)\vec{x}, \label{eq:power_iteration_1}\\
    \vec{x} &\leftarrow \vec{x} / \|\vec{x}\|_2.
\end{align}
Here, matrix $\mat{S}$ is a shift operator. For now, we assume that $\mat{S}=0$ and will revisit $\mat{S}$ shortly.
Equation~\eqref{eq:power_iteration_1} involves the product of dense modularity matrix $\mat{Q}$ and dense vector $\vec{x}$, which scales with ${\cal O}(n^2)$, where $n$ is the number of nodes.

The power-iteration method leverages the fact that the modularity matrix is the sum of a sparse matrix, $\mat{A}$, and a dense matrix, $\frac{\vec{k}\vec{k}^\top}{2m}$. This allows us to rewrite Eq.~\eqref{eq:power_iteration_1} as
\begin{align}
    \label{eq:power_iteration_2}
    \vec{x} \leftarrow \frac{1}{2m}\left(\mat{A}\vec{x} - \frac{(\vec{k}^\top \vec{x})^2}{2m}\right) - \mat{S}\vec{x}
\end{align}
Here, $\mat{A}\vec{x}$ has a complexity of ${\cal O}(m)$, and $\vec{k}^\top \vec{x}$ has a complexity of ${\cal O}(n)$. Thus, the total computational complexity of Eq.~\eqref{eq:power_iteration_2} is ${\cal O}(m+n)$, which is significantly less than the ${\cal O}(n^2)$ complexity of Eq.~\eqref{eq:power_iteration_1}.

Now, let's revisit $\mat{S}$. This matrix is useful when computing the secondary eigenvectors. Let $\vec{x}^{\ell}$ be the $\ell$th eigenvector associated with the $\ell$th largest eigenvalue in magnitude.
Let us consider that we have already computed the eigenvectors $\vec{x}^{1}, \vec{x}^{2}, \ldots, \vec{x}^{\ell-1}$ and their corresponding eigenvalues $\lambda_1, \lambda_2, \ldots, \lambda_{\ell-1}$.
If we could compute $\vec{x}^{\ell}$, then we would be able to compute the modularity embedding by computing the eigenvectors $\vec{x}^{1}, \vec{x}^{2}, \ldots, \vec{x}^{n}$ sequentially.
To compute $\vec{x}^{\ell}$, we remove the contribution of the eigenvectors $\vec{x}^{1}, \vec{x}^{2}, \ldots, \vec{x}^{\ell-1}$ from the modularity matrix.
Matrix $\mat{S}$ represents these contributions.
Specifically, assuming that the network is undirected and thereby the left and right eigenvectors are equal, then the modularity matrix is decomposed by
\begin{align}
    Q = \sum_{\ell=1} ^n \lambda_{\ell}\vec{x}^{\ell}(\vec{x}^{\ell})^\top
\end{align}
where $\lambda_\ell$ is the eigenvalue of the eigenvector $\vec{x}^{\ell}$.
Thus, the contribution of $\vec{x}^{1}, \vec{x}^{2}, \ldots, \vec{x}^{\ell-1}$ for $\mat{Q}$ is given by
\begin{align}
    \mat{S} = \sum_{s =1}^{\ell-1} \lambda_s \vec{x}^{s}(\vec{x}^{s})^\top,
\end{align}
By plugging $\mat{S}$ into Eq.~\eqref{eq:power_iteration_2}, we obtain
\begin{align}
    \label{eq:power_iteration_3}
    \vec{x}^{\ell} \leftarrow \frac{1}{2m}\left(\mat{A}\vec{x}^{\ell} - \frac{(\vec{k}^\top \vec{x}^{\ell})^2}{2m}\right) -
    \sum_{s =1}^{\ell-1} \lambda_s \left( (\vec{x}^{s})^\top \vec{x}^{\ell} \right)\vec{x}^{s}.
\end{align}
Equation~\eqref{eq:power_iteration_3} involves additional $\ell-1$ vector-by-vector product terms, which scales with ${\cal O}(\ell n)$, resulting in the total complexity of ${\cal O}(m + \ell n)$ for computing $\ell$th eigenvector.
This power iteration method scales with the network size and is memory efficient because it does not require the construction of the dense modularity matrix.

%
%
%
\subsection{Non-backtracking walk embedding}

Following~\cite{krzakalaSpectralRedemptionClustering2013}, we computed the eigenvectors of the non-backtracking matrix. We use \texttt{scipy.linalg.eigs} implemented in the scipy package~\cite{2020SciPy-NMeth}. Because the solver did not converge for some networks, we relaxed the convergence criterion by setting \texttt{tol=0.0001}.

\subsection{Flat SBM}

We used the degree-corrected stochastic block model without hierarchical partitioning implemented in the \texttt{graph-tool} package~\cite{GraphtoolEfficentNetwork}. Since we focus on the basic clustering ability of the method, the number of communities is set to the number of true communities.

\subsection{Belief propagation}

Belief propagation is an optimal method for sparse networks generated by the stochastic block model. We employed the code, \texttt{sbm}, provided by an author of the original paper~\cite{zhangScalableDetectionStatistically2014,itpZhangapossWebsite}. We set all the parameters based on the true communities.
More specifically, we set the number of communities to the true number of communities. We then set the ``cab matrix''---specified by the \texttt{-c} option of \texttt{sbm}---to the density of edges between and within groups, multiplied by the number of nodes. Lastly, we set the fractional group size specified by the \texttt{-P} option of \texttt{sbm} to the fraction of nodes in each true community. We made available our Python wrapper for \texttt{sbm} at~\cite{githubGitHubSkojakuBeliefPropagation}.

\section{Embedding with a smaller number of dimensions}

\begin{figure}[h!]
\centering
\includegraphics[width=\hsize]{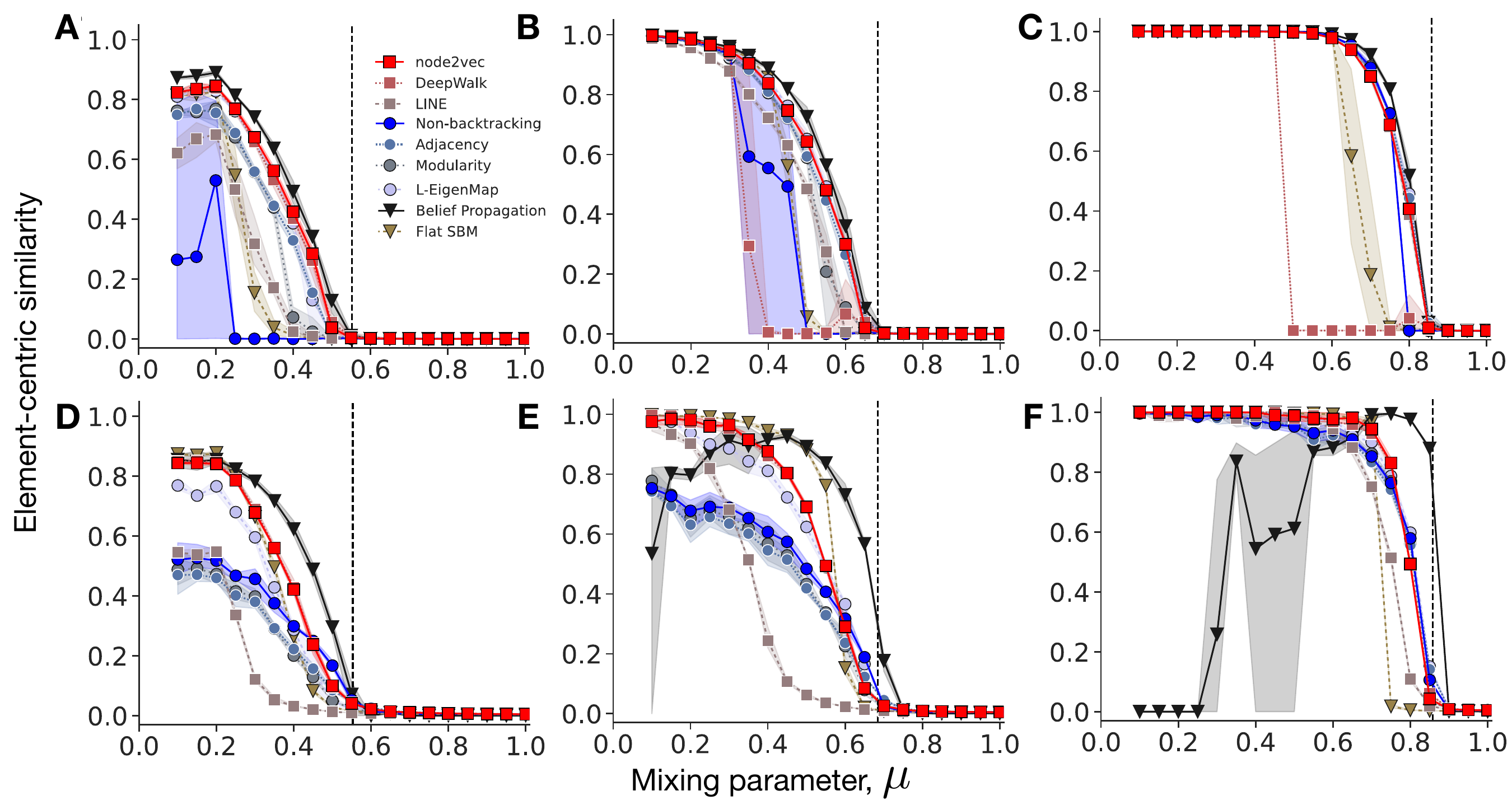}
\caption{
Performance plots for the planted partition model. The number of dimensions of all embeddings is $C=16$.
We generated networks with $n = 10^5$ nodes, different edge sparsity ($\langle k \rangle=5$ in A and D, $\langle k \rangle=10$ in B and E, $\langle k \rangle=50$ in C and F), and the different number of communities ($q=2$ for A--C and $q=50$ for D--F).
The dashed vertical line indicates the theoretical detectability limit $\mu^*$. 
}
\label{fig:multipartiton_model_dim16}
\end{figure}

\begin{figure}[h!]
\centering
\includegraphics[width=\hsize]{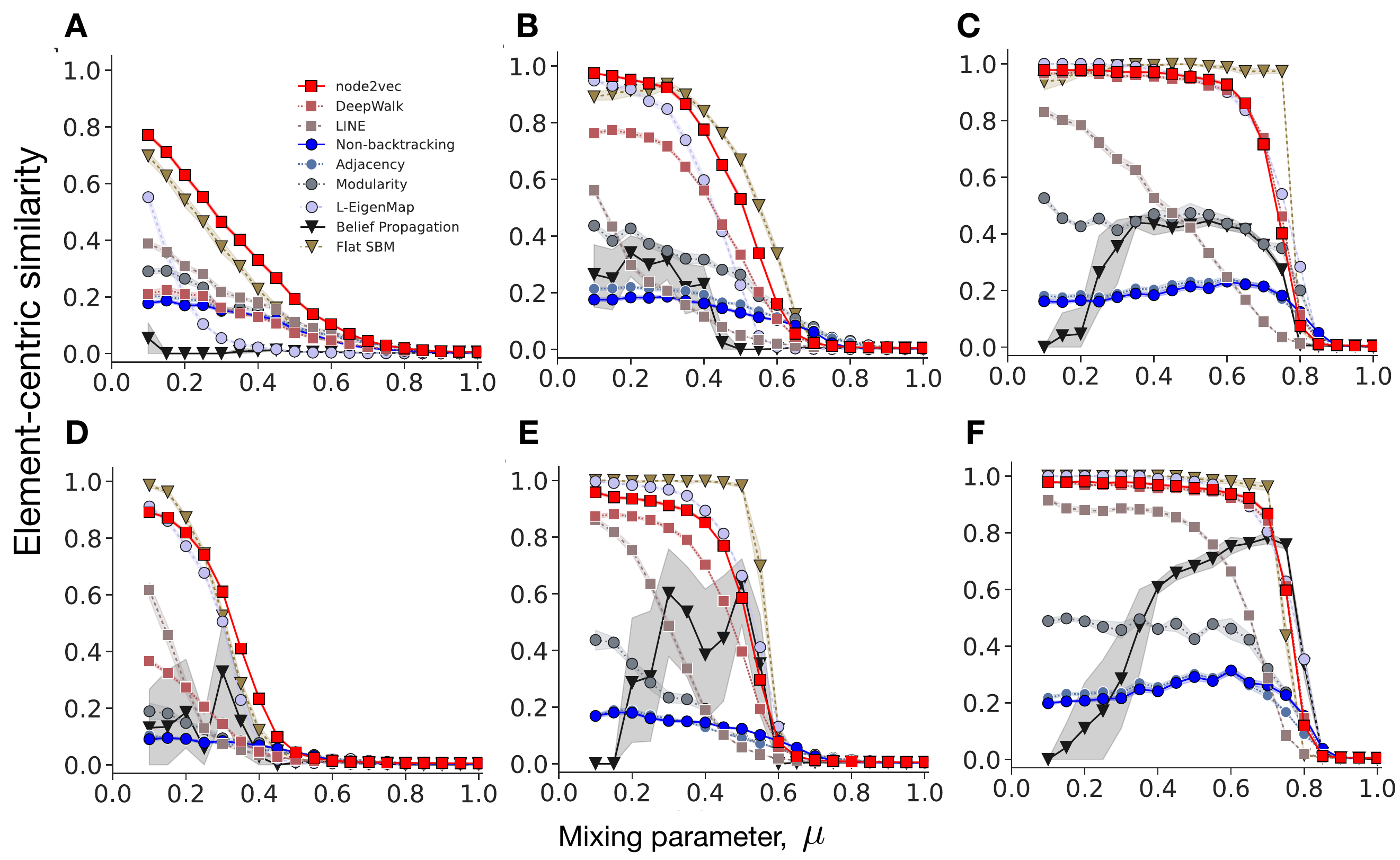}
\caption{
Performance plots for the LFR benchmark. The number of dimensions of all embeddings is $C=16$.
We generated networks with $n = 10^4$ nodes with different edge sparsity ($\langle k \rangle=5$ in A and D, $\langle k \rangle=10$ in B and E, $\langle k \rangle=50$ in C and F).
The degree exponent $\tau_1=2.1$ in A, B, and C, and $\tau_1=3$ in D, E, and F.
}
\label{fig:lfr_dim16}
\end{figure}

We tested graph embeddings with a smaller number of dimensions, i.e., $C=16$, usign the networks of $10,000$ nodes generated by the PPM (Figs.~\ref{fig:multipartiton_model_dim16} and \ref{fig:lfr_dim16}).
Despite the fact that $C$ is lower than the number of true communities $q$ in the examples where $q=50$, we find qualitatively the same results as in the results for $C=64$, that we reported in the main text.

Specifically, for the stochastic block model, the community detection performance decreases compared to $C=64$ dimensions, and node2vec outperforms other graph embedding methods for most values of the mixing parameter $\mu$.
The performance of node2vec stands out for $q=50$, where the number of communities is larger than the number of dimensions (Figs.~\ref{fig:multipartiton_model_dim16}D--F).
For the LFR model, the community detection performance decreases overall (Fig.~\ref{fig:lfr_dim16}). Nevertheless, node2vec is comparable to or outperforms Infomap.

\section{Robustness of the node2vec performance}
\label{sec:robustness}
node2vec has several hyperparameters that may impact its performance. Thus, we explored a broad range of hyperparameters to examine the robustness of their performance for the PPM and LFR networks by using the K-means clustering.

\begin{figure}
  \includegraphics[width=\hsize]{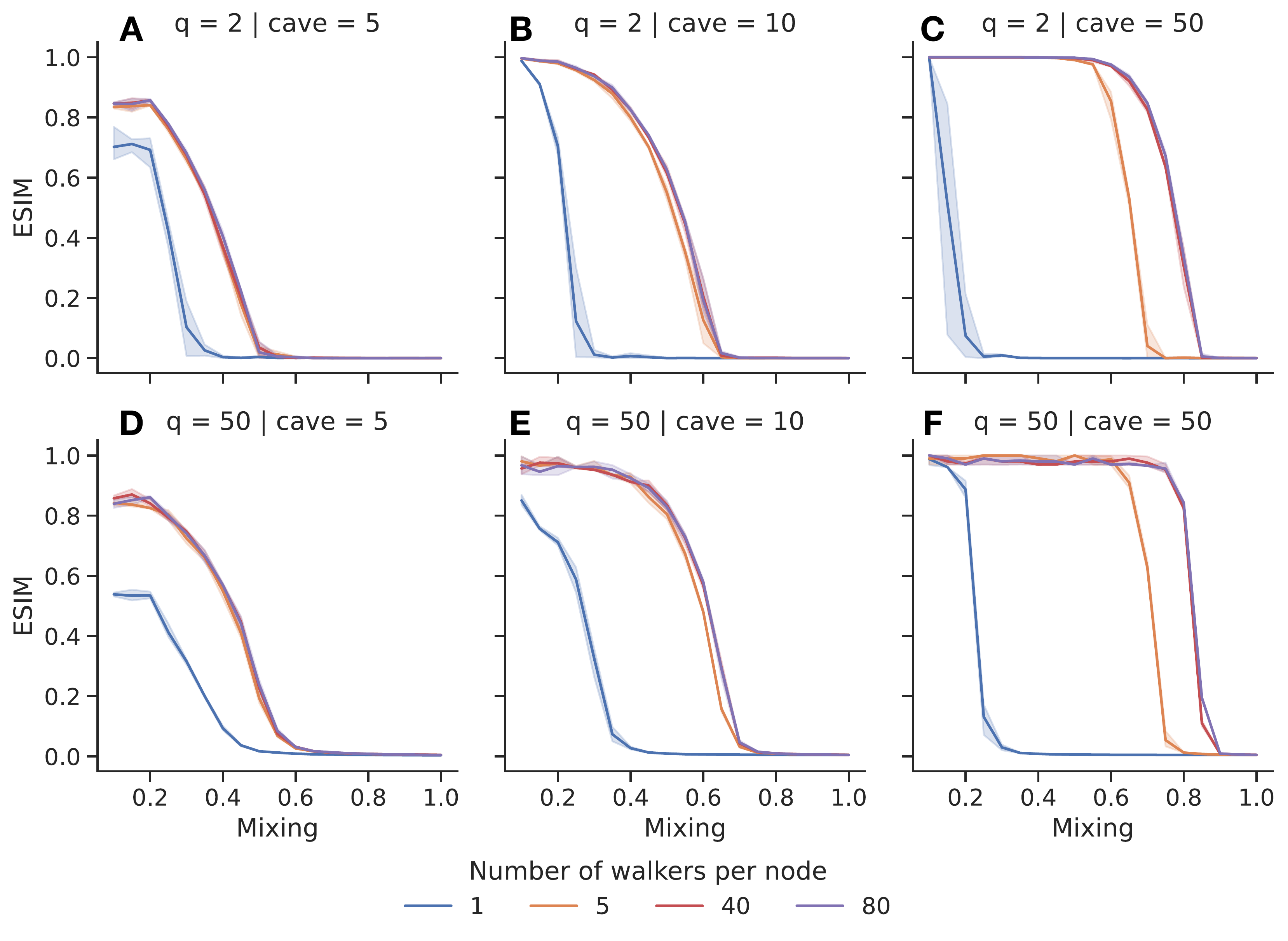}
  \caption{%
  Community detection performance of node2vec with different values of \emph{random walkers per node} for the PPM networks.
  We generated networks with $n=10^4$ nodes, different edge sparsity ($\langle k\rangle$ = 5 in A and D, $\langle k\rangle$ = 10 in B and E, and $\langle k\rangle$ = 50 in C and F), and
  different number of communities ($q=2$ for A--C and $q=50$ for D--F).
  The dashed vertical line indicates the theoretical detectability limit $\mu^*$.
  The black line represents the belief propagation method. The other colored lines represent node2vec, with different hyperparameter values (i.e., number of random walkers per node).
  }
  \label{fig:mpl_var=nWalks}
  \end{figure}

\begin{figure}
  \includegraphics[width=\hsize]{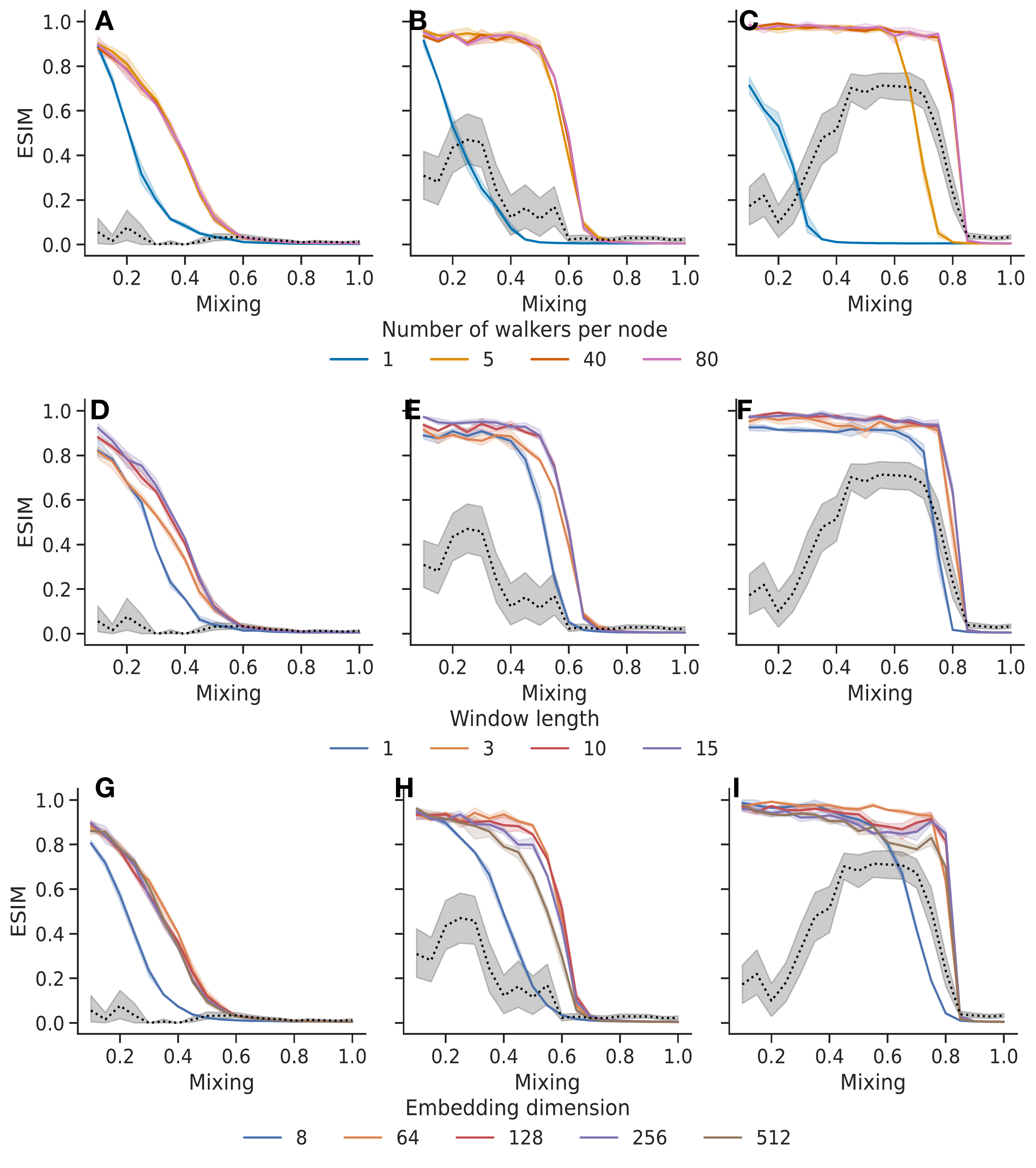}
  \caption{%
  Community detection performance of node2vec for the LFR network with $10^4$ nodes with different edge sparsity ($\langle k\rangle$ = 5 in A, D, G, $\langle k\rangle$ = 10 in B, E, H,  and $\langle k\rangle$ = 50 in C, F, I).
  The black line represents the belief propagation method.
  The other colored lines represent node2vec, with different hyperparameter values (i.e., embedding dimensions).
  The degree exponent is set to $\tau_1 = 3.0$.
  }
  \label{fig:lfr_robustness}
\end{figure}

The number of random walkers per node is a key hyperparameter for node2vec.
In node2vec, random walkers traverse the network to generate sequences of visited nodes. Multiple walkers are placed at each node to produce multiple sequences to train the node2vec model. The number of walkers determines the amount of training data and thus impacts performance.
We found that using only 1 walker performs poorly due to insufficient training data (Figs.~\ref{fig:mpl_var=nWalks} and \ref{fig:lfr_robustness}A--C). However, using 5 or more walkers gives a comparable performance, indicating that node2vec is robust above a minimum threshold of walkers needed to gather enough training data.

\begin{figure}
  \includegraphics[width=\hsize]{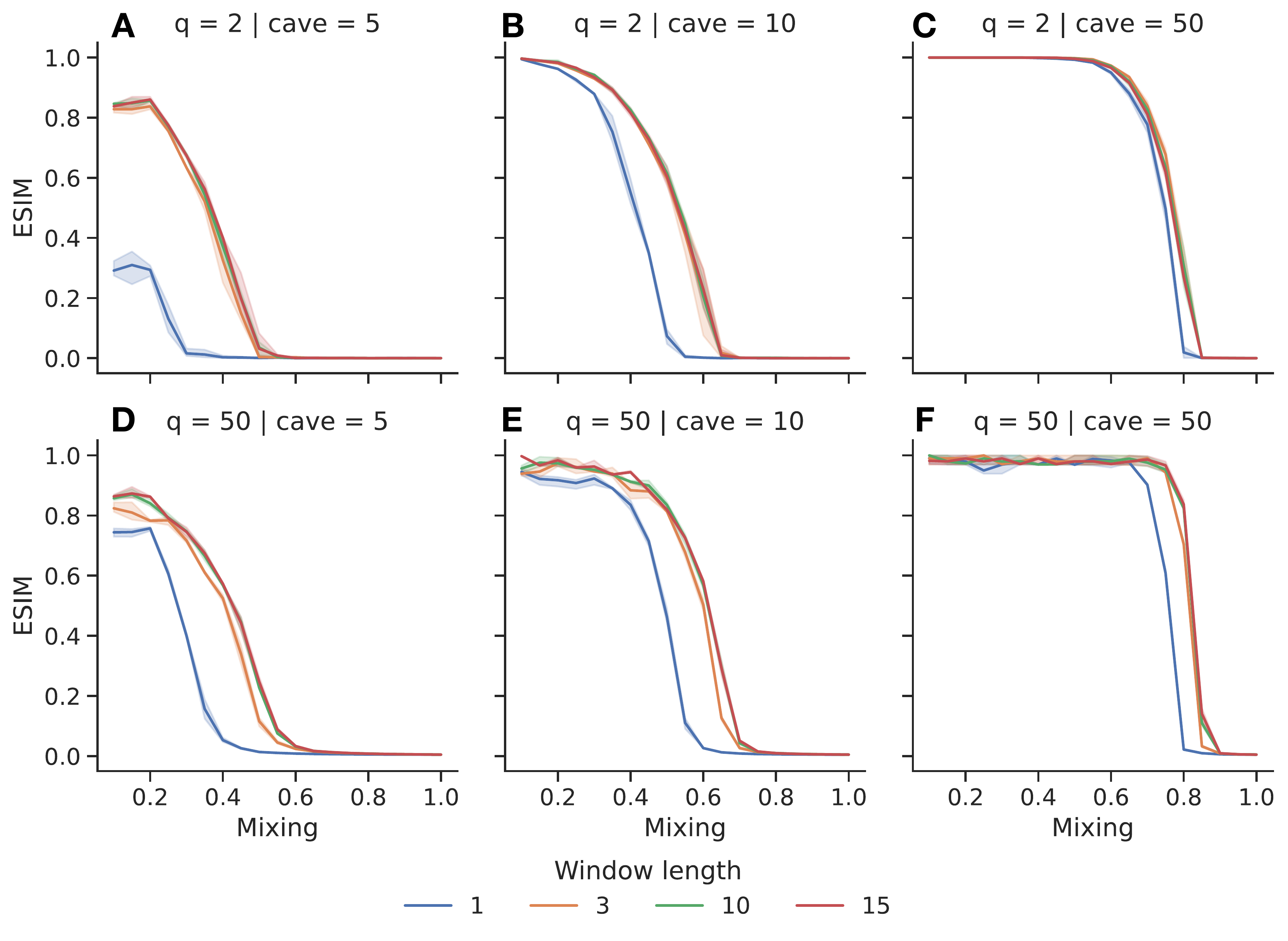}
  \caption{%
  Community detection performance of node2vec with different values of \emph{window length} for the PPM networks.
  We generated networks with $n=10^4$ nodes, different edge sparsity ($\langle k\rangle$ = 5 in A and D, $\langle k\rangle$ = 10 in B and E, and $\langle k\rangle$ = 50 in C and F), and
  different number of communities ($q=2$ for A--C and $q=50$ for D--F).
  The dashed vertical line indicates the theoretical detectability limit $\mu^*$.
  The black line represents the belief propagation method. The other colored lines represent node2vec, with different hyperparameter values (i.e., window length).
  }
  \label{fig:mpl_var=windowLength}
  \end{figure}

The window size is another key parameter controlling the scale of the network structure encoded in the embedding. Smaller windows emphasize local structure, while larger windows capture more global properties. We found community detection performance increases with window size up to $L=10$ (Fig.~\ref{fig:mpl_var=windowLength} and \ref{fig:lfr_robustness}D--F), but larger values result in marginal substantial performance gain.

\begin{figure}
  \includegraphics[width=\hsize]{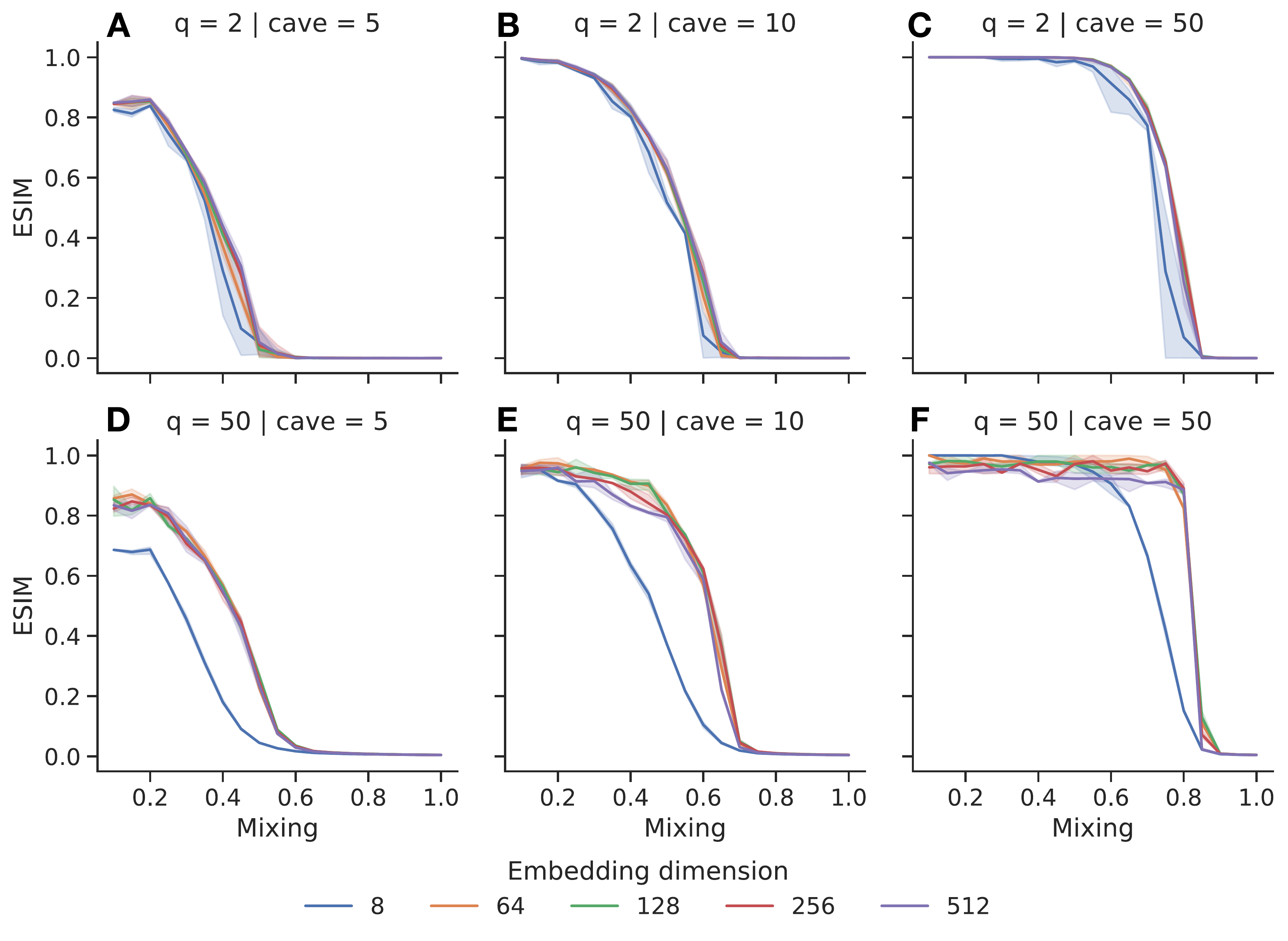}
  \caption{%
  The community detection performance of node2vec with different values of \emph{embedding dimensions} for the PPM networks.
  We generated networks with $n=10^4$ nodes, different edge sparsity ($\langle k\rangle$ = 5 in A and D, $\langle k\rangle$ = 10 in B and E, and $\langle k\rangle$ = 50 in C and F), and
  different number of communities ($q=2$ for A--C and $q=50$ for D--F).
  The dashed vertical line indicates the theoretical detectability limit $\mu^*$.
  The black line represents the belief propagation method. The other colored lines represent node2vec, with different hyperparameter values (i.e., the number of embedding dimensions).
  }
  \label{fig:mpl_var=dim}
  \end{figure}

Finally, we examined the embedding dimensionality (Figs.~\ref{fig:mpl_var=dim} and \ref{fig:lfr_robustness}G--I). Higher dimensions improve performance, especially for sparse networks.
Our choice of 64 dimensions is suboptimal, although it was still sufficient for node2vec to compete with or exceed all other methods. This indicates the potential for even better performance, given more dimensionality.

In summary, we confirmed that node2vec is robust across a range of hyperparameters. The original parameter choices provide good performance, and there is great potential for even greater performance by optimizing them for specific networks.

\section{Embedding with the Voronoi clustering algorithm}

The performance of community detection depends on both the quality of the embedding and the performance of the subsequent data clustering procedure.
While we used the $K$-means algorithm to identify the clusters, it is possible that the $K$-means algorithm substantially degrades the performance of community detection even when the embedding captures the community structure.
To eliminate potential noises from the clustering step, we tested a variant of the $K$-means algorithm with fixed centroids (i.e., Voronoi clustering), whose positions are determined by the locations of the true communities in the embedding space. Clustering is performed by assigning each point/node to the centroid/cluster with the highest cosine similarity.
Using this algorithm, we can focus on whether an embedding method can successfully \textit{encode} community structure.

\begin{figure}[h!]
\centering
\includegraphics[width=\hsize]{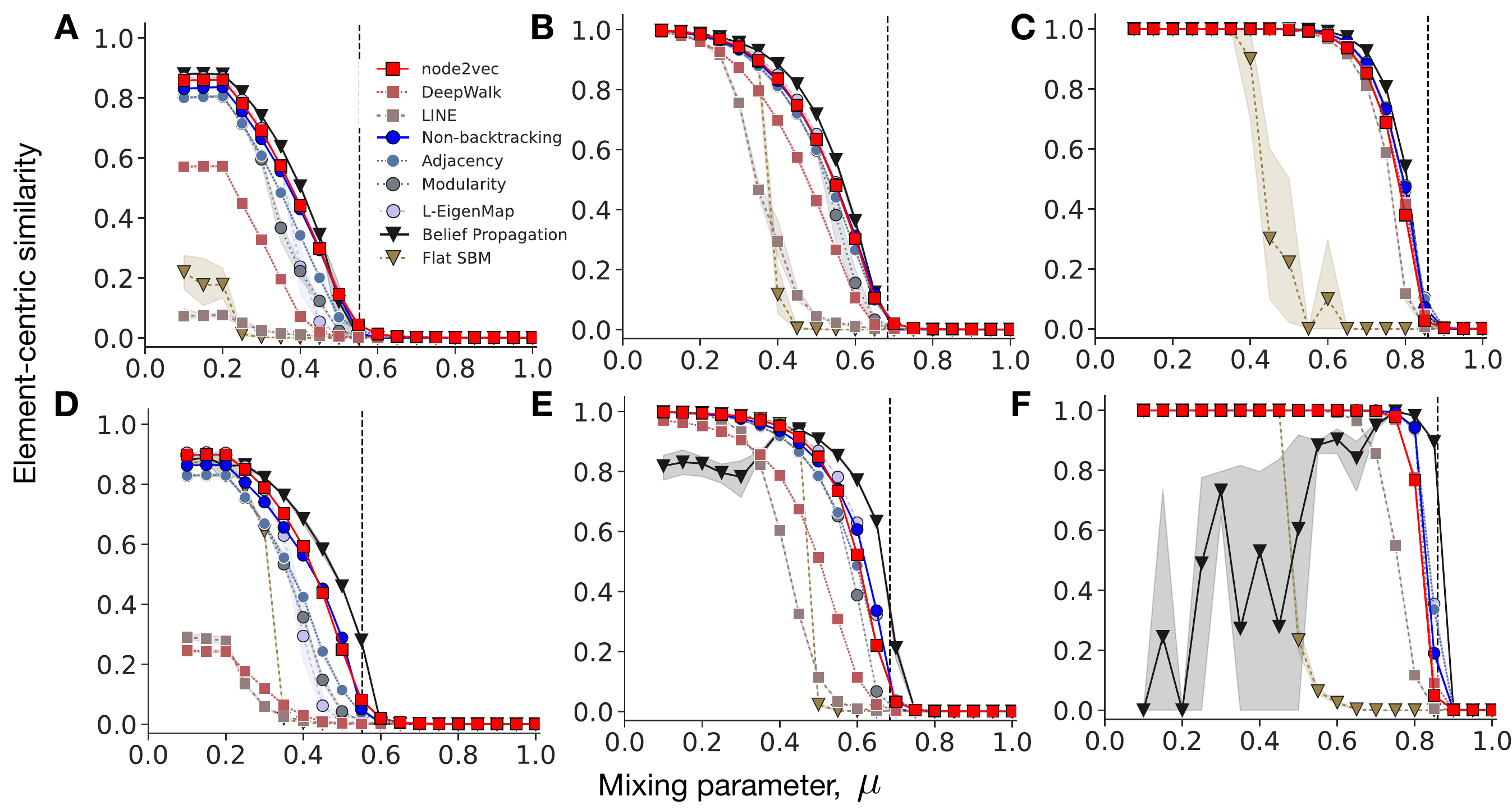}
\caption{
Performance plots for the planted partition model. Here we used the Voronoi clustering to identify the communities in the embedding. The number of dimensions is $C=64$.
We generated networks with $n = 10^5$ nodes, different edge sparsity ($\langle k \rangle=5$ in A and D, $\langle k \rangle=10$ in B and E, $\langle k \rangle=50$ in C and F), and the different number of communities ($q=2$ for A--C and $q=50$ for D--F).
}
\label{fig:multipartition_model_voronoi}
\end{figure}

\begin{figure}[h!]
\centering
\includegraphics[width=\hsize]{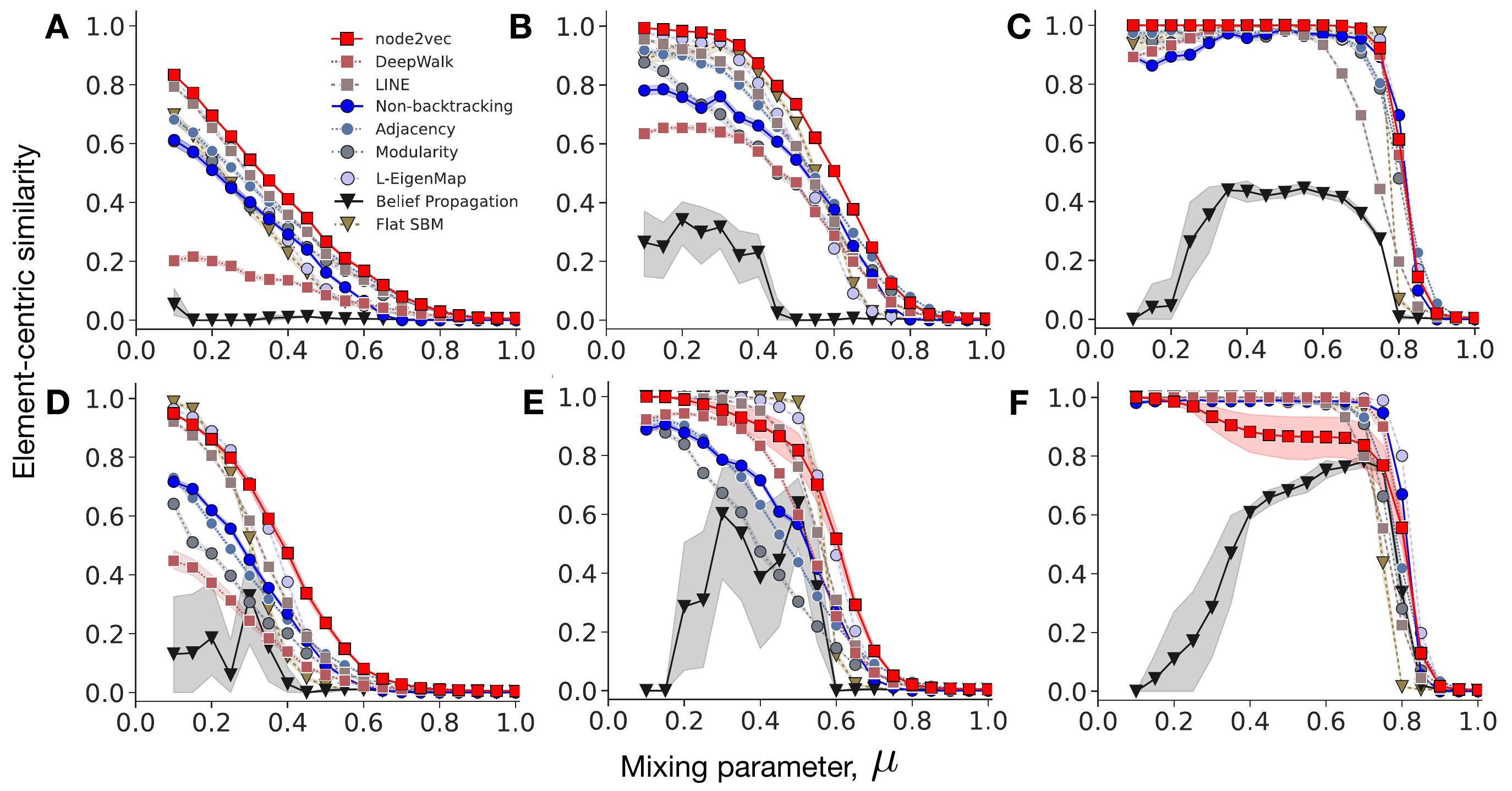}
\caption{%
Performance plots for the LFR benchmark. Here we used the Voronoi clustering to identify the communities in the embedding. The number of dimensions is $C=64$.
We generated networks with $n = 10^4$ nodes with different edge sparsity ($\langle k \rangle=5$ in A and D, $\langle k \rangle=10$ in B and E, $\langle k \rangle=50$ in C and F).
The degree exponent $\tau_1=2.1$ in A, B, and C, and $\tau_1=3$ in D, E, and F.
}
\label{fig:lfr_voronoi}
\end{figure}

The Voronoi clustering algorithm yields a comparable performance with the $K$-means algorithm for the PPM networks (Fig.~\ref{fig:multipartition_model_voronoi}), suggesting that the $K$-means algorithm is nearly optimal for the embeddings generated by the methods we tested for the PPM networks.
For the LFR networks, the Voronoi clustering algorithm yielded noticeably higher performance, especially for the networks with small $\mu$ values (Fig.~\ref{fig:lfr_voronoi}).
The performance improvement is substantial for node2vec, reaching the best or on par with the top-performing method.
The results suggest that the clustering algorithm can largely limit the effectiveness of embedding in the community detection task.

\clearpage


\clearpage

\renewcommand{\refname}{Supplementary References}